\documentclass[aps,prb,twocolumn,groupedaddress,draft,intlimits,amsmath,amssymb,floats]{revtex4}

\usepackage{bm}
\usepackage[final]{graphicx}
\usepackage{epsfig}
\usepackage{subfigure}
\emergencystretch=\maxdimen
\begin{document}

\title{Characterizing the three-orbital Hubbard model with determinant quantum Monte Carlo} 
\author{Y. F. Kung$^{1,2}$}
\author{C.-C. Chen$^{3,4}$}
\author{Yao Wang$^{2,5}$}
\author{E.W. Huang$^{1,2}$}
\author{E. A. Nowadnick$^{1,2,6}$}
\author{B. Moritz$^{2,7}$}
\author{R. T. Scalettar$^8$}
\author{S. Johnston$^{9}$}
\author{T. P. Devereaux$^{2,10}$}

\affiliation{$^1$Department of Physics, Stanford University, Stanford, CA 94305 USA}
\affiliation{$^2$Stanford Institute for Materials and Energy Sciences, SLAC National Accelerator Laboratory and Stanford University, Stanford, CA 94305, USA}
\affiliation{$^3$Advanced Photon Source, Argonne National Laboratory, Lemont, Illinois 60439, USA}
\affiliation{$^4$Department of Physics, University of Alabama at Birmingham, Birmingham, Alabama 35294, USA}
\affiliation{$^5$Department of Applied Physics, Stanford University, California 94305, USA}
\affiliation{$^6$School of Applied and Engineering Physics, Cornell University, Ithaca, NY 14853 USA}
\affiliation{$^7$Department of Physics and Astrophysics, University of North Dakota, Grand Forks, ND 58202, USA}
\affiliation{$^8$Department of Physics, University of California - Davis, Davis, CA 95616, USA}
\affiliation{$^9$Department of Physics and Astronomy, University of Tennessee, Knoxville, Tennessee 37996, USA}
\affiliation{$^10$Geballe Laboratory for Advanced Materials, Stanford University, Stanford, CA 94305, USA}

\begin{abstract}
We characterize the three-orbital Hubbard model using state-of-the-art determinant quantum Monte Carlo (DQMC) simulations with parameters relevant to the cuprate high-temperature superconductors.  The simulations find that doped holes preferentially reside on oxygen orbitals and that the $(\pi,\pi)$ antiferromagnetic ordering vector dominates in the vicinity of the undoped system, as known from experiments.  The orbitally-resolved spectral functions agree well with photoemission spectroscopy studies and enable identification of orbital content in the bands.  A comparison of DQMC results with exact diagonalization and cluster perturbation theory studies elucidates how these different numerical techniques complement one another to produce a more complete understanding of the model and the cuprates.  Interestingly, our DQMC simulations predict a charge-transfer gap that is significantly smaller than the direct (optical) gap measured in experiment.  Most likely, it corresponds to the indirect gap that has recently been suggested to be on the order of 0.8 eV, and demonstrates the subtlety in identifying charge gaps.
\end{abstract}


\maketitle

\section{Introduction}
The copper-oxide planes of the cuprate superconductors are believed to support a rich variety of phases, such as high-temperature superconductivity, magnetism, and charge-stripe order, as well as cooperation and competition between them.  As a result, most theoretical work focuses on describing the planes, the first step of which is to determine how many and which of the copper and oxygen orbitals to include in a minimal model.  One of the most commonly used is the single-band Hubbard model,\cite{Hubbard_PRSLA_1963,Anderson_PR_1959} which captures experimental features such as antiferromagnetism in the undoped and lightly doped compounds.\cite{Dagotto_RMP_1994}  However, despite these successes, it is a low-energy effective model that assumes that the oxygen degrees of freedom do not contribute significantly to the physics and that the quasiparticles are Zhang-Rice singlets (ZRS).\cite{Zhang_PRB_1988}  As such, it is not entirely clear how accurately it captures cuprate physics, especially at higher energies and for proposed states that explicitly involve the oxygen orbitals.\cite{Varma_PRB_1997, Varma_PRL_1999,Varma_PRB_2006, Varma_PRB_2012}  As experiments have determined that doped holes preferentially reside on oxygen orbitals, it has been argued that including oxygen explicitly is crucial to understanding the physics; indeed, recent calculations have found that a doped hole moves on the oxygen sublattice and that its dynamics are relatively unaffected by spin fluctuations on copper.\cite{Sawatzky_PRL_2011,Sawatzky_NatPhys_2014}

The three-orbital Hubbard model provides a more realistic picture of the copper oxide planes, as it includes the copper $3d_{x^2-y^2}$ orbitals as well as the neighboring oxygen $2p_x$ and $2p_y$ orbitals.\cite{Mattheiss_PRL_1987,Varma_SSC_1987, Emery_PRL_1987}  To assess its accuracy in describing the cuprates, quantities such as spectral functions can be calculated and compared to experiments, including optical conductivity measurements, angle-resolved photoemission spectroscopy (ARPES), O $K$-edge x-ray absorption spectroscopy, and Cu $K$- and $L$-edge resonant inelastic x-ray scattering spectroscopy.\cite{Dagotto_RMP_1994,Weber_PRB_2008,deMedici_PRB_2009,Chen_PRL_2010,Wang_PRB_2010,Chen_PRB_2013,Go_PRL_2015}  The explicit inclusion of oxygen orbitals enables a proper description of the ZRS in order to determine whether the ZRS picture is still applicable at high doping levels,\cite{Chen_PRB_2013} as well as a systematic evaluation of various proposals for the pseudogap regime, such as oxygen antiferromagnetism (AFM) and orbital loop currents that circulate between copper and oxygen orbitals.\cite{Kung_PRB_2014}  The model can thus address the issue of when it may be necessary to include oxygen in order to model the cuprates.

In general, the three-orbital Hubbard model is too complicated to solve analytically, so numerous computational methods have been brought to bear on the problem, such as determinant quantum Monte Carlo (DQMC),\cite{BSS_PRD_1981,White_PRB_1989,Dopf_PRB_1990,Scalettar_PRB_1991,Kuroki_PRL_1996} exact diagonalization (ED),\cite{Dagotto_RMP_1994,ARPACK,Chen_PRL_2010} cluster perturbation theory (CPT),\cite{Senechal_PRL_2000,Senechal_PRB_2002} density matrix renormalization group (DMRG),\cite{Schollwoeck_RMP_2005,Verstraete_AdvPhys_2008,Schollwoeck_AnPhys_2011,White_PRB_2015} and dynamical mean field theory (DMFT).\cite{Georges_RMP_1996,Go_PRL_2015}  DQMC and ED both have the advantage of being numerically exact.  DQMC (discussed in detail in the next section) can treat large system sizes, but is restricted to high temperatures due to the fermion sign problem.  On the other hand, ED solves the eigenvalue problem for energies and wavefunctions of the Hamiltonian, using iterative Krylov subspace methods.  In general, it is performed at zero-temperature, but its main drawback is that the number of states in the Hilbert space grows rapidly with system size, limiting simulations to relatively small clusters (current state-of-the-art ED calculations have studied Cu$_8$O$_{16}$ planar clusters from the undoped to overdoped regime\cite{Chen_PRL_2010,Chen_PRB_2013}).  CPT combines ED and perturbation theory, dividing the infinite plane into smaller identical clusters that are solved exactly using ED.  Hopping between the clusters is treated to leading order in perturbation theory.  CPT is exact in the limits of strong and weak coupling as the number of Brillouin zone sites $L\rightarrow \infty$ and, like ED, is generally performed at zero-temperature.  It offers the advantages of fine momentum resolution in the spectral function and a better approximation of the thermodynamic limit than ED; however, unlike DQMC and ED, it is restricted to calculating single-particle quantities.  

Other techniques that have been used to study the three-orbital model are DMRG and DMFT.  DMRG accesses larger system sizes than ED by truncating the Hilbert space to keep only the most significant basis functions.  Because it does not suffer from the fermion sign problem, it can also reach lower temperatures than DQMC.  However, unlike DQMC, it is limited to quasi-one-dimensional systems, although there are efforts to extend it and to develop analogues in order to study higher dimensional systems.\cite{Verstraete_AdvPhys_2008}  A recent DMRG study has explored the spin and charge structures of hole- and electron-doped systems.\cite{White_PRB_2015}  DMFT, on the other hand, maps the many-body interacting problem to an impurity model embedded in a mean field and assumes a local lattice self-energy.  However, its accuracy is limited by how well the local picture can capture the self-energy.  It has been used to map the phase diagram of the three-orbital model, but the underdoped cuprates, which exhibit interesting phenomena such as the pseudogap regime and Fermi arcs, has a non-local self-energy, thus limiting the usefulness of DMFT.  Recently, the method has been extended to embed small clusters in the mean field to capture some of this momentum dependence.\cite{Kent_PRB_2008,Weber_EPL_2012,Go_PRL_2015}  Efforts have been made to treat even larger systems by using DMRG as the impurity solver for DMFT.\cite{Wolf_PRX_2015}

In this paper, we use DQMC to characterize properties of the three-orbital Hubbard model because it is a numerically exact method that accesses larger, two-dimensional systems and captures non-local effects.  We perform calculations on square systems up to Cu$_{36}$O$_{72}$ ($N=36$) in size and with temperatures down to $\beta=30$ eV$^{-1}$, and compare the results to complementary ones obtained from CPT and ED calculations.  The paper is organized as follows: Section II reviews the three-orbital Hubbard model and DQMC algorithm.  Section III explores the effects of different parameters on the fermion sign, while Section IV examines the doping dependence of the potential and kinetic energies and static correlations, commenting on connections to experiment.  Section V presents the orbitally-resolved spectral functions and density of states, which can be compared to experimental results and other theoretical approaches.  In Section VI, DQMC results are complemented by those from ED and CPT to form a more complete picture of the model.  Finally, we offer some concluding remarks in Section VII.

\section{Model and methods}
\subsection{Three-orbital Hubbard model}

\begin{figure}[b!]
	\includegraphics[scale=0.3]{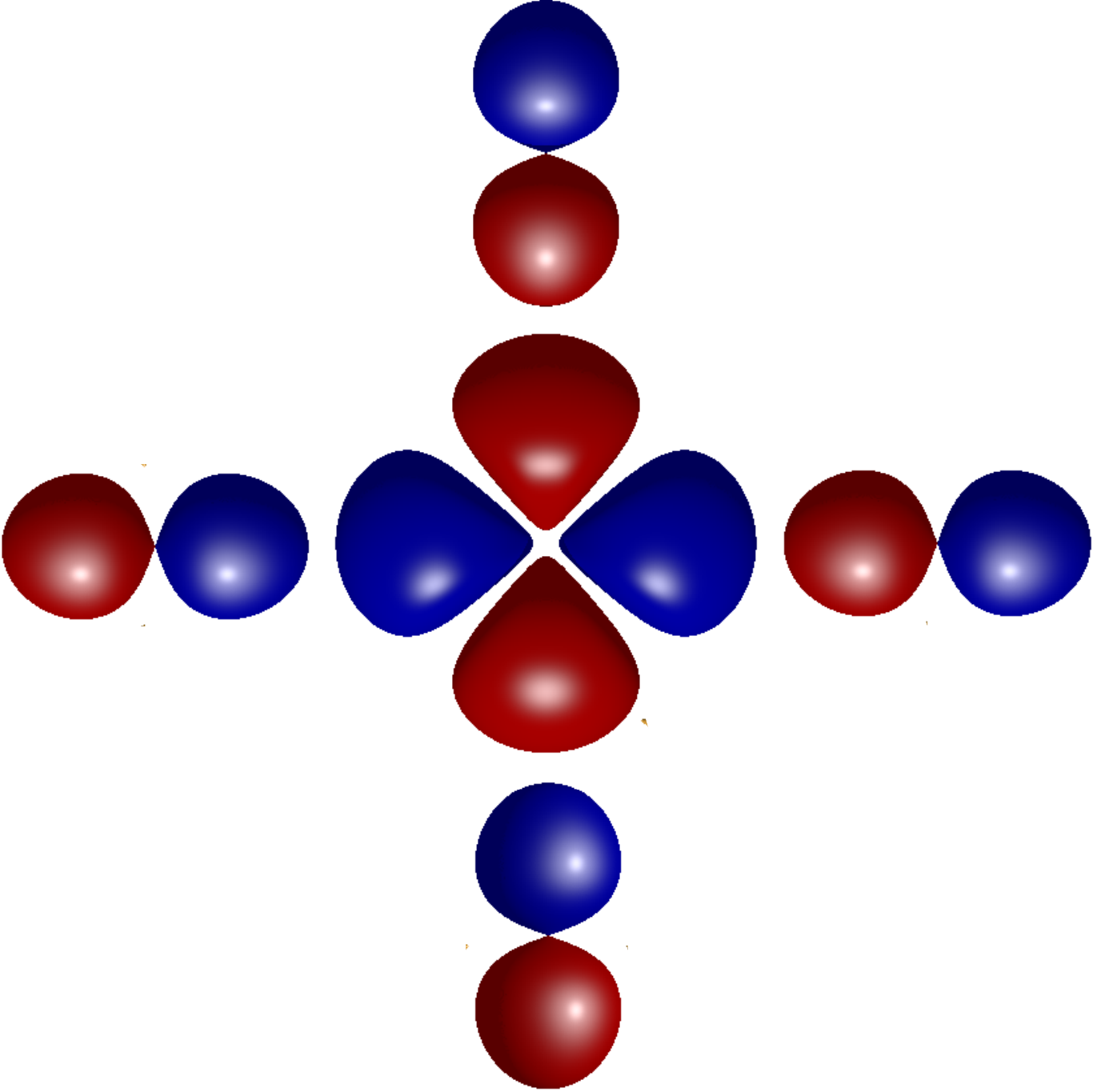}
	\caption{A copper $d_{x^2-y^2}$ orbital and its surrounding oxygen $p_x$ or $p_y$ orbitals are shown.  The colors indicate the phase factors (blue for positive, red for negative).}
	\label{fig:cartoon}
\end{figure}

The three-orbital Hubbard model treats the copper-oxide planes of the cuprates, with copper $3d_{x^2-y^2}$ orbitals and oxygen $2p_x$ and $2p_y$ orbitals described by the Hamiltonian $H=K_{pd} + K_{pp} + V_{dd} + V_{pp}$.\cite{Mattheiss_PRL_1987,Varma_SSC_1987, Emery_PRL_1987}  The kinetic energy terms $K_{pd}$ and $K_{pp}$ describe the copper-oxygen and oxygen-oxygen hopping, respectively, as

\begin{eqnarray}
K_{pd}&=&\sum_{\langle i,j \rangle \sigma} (t_{ij}^{\phantom{\dagger}} d^\dagger_{i\sigma} c_{j\sigma}^{\phantom{\dagger}} + h.c.)\nonumber\\
&=&\sum_{\langle i,j \rangle \sigma} d^\dagger_{i\sigma} k_{i,j}^{pd} c^{\phantom{\dagger}}_{j\sigma},\nonumber\\
K_{pp}&=&\sum_{\langle j,j^\prime \rangle \sigma} (t_{jj^\prime}^{\phantom{\dagger}} c_{j\sigma}^\dagger c_{j^\prime \sigma}^{\phantom{\dagger}} + h.c.)\nonumber\\
&=&\sum_{\langle j,j^\prime \rangle \sigma} c^\dagger_{j\sigma} k_{j,j^\prime}^{pp} c^{\phantom{\dagger}}_{j^\prime\sigma},\nonumber\\
\end{eqnarray}
where $d^\dagger_{i\sigma}$ ($d_{i\sigma}^{\phantom{\dagger}}$) creates (destroys) a hole with spin $\sigma$ on a copper orbital at site $i$, $c_{j\sigma}^\dagger$ ($c_{j\sigma}^{\phantom\dagger}$) creates (destroys) a hole with spin $\sigma$ on an oxygen orbital at site $j$, and
\begin{eqnarray}
t_{ij}&=&t_{pd}(-1)^{\eta_{ij}},\nonumber\\
t_{jj^\prime}&=&t_{pp}(-1)^{\beta_{jj^\prime}}.
\end{eqnarray}
In hole language, the phase convention is $\eta_{ij}=1$ for $j=i+\frac{1}{2}\hat{x}$ or $j=i-\frac{1}{2}\hat{y}$ and $\eta_{ij}=0$ for $j=i-\frac{1}{2}\hat{x}$ or $j=i+\frac{1}{2}\hat{y}$.  In addition, $\beta_{jj^\prime}=1$ for $j^\prime=j-\frac{1}{2}\hat{x}-\frac{1}{2}\hat{y}$ or $j^\prime=j+\frac{1}{2}\hat{x}+\frac{1}{2}\hat{y}$ and $\beta_{jj^\prime}=0$ for $j^\prime=j-\frac{1}{2}\hat{x}+\frac{1}{2}\hat{y}$ or $j^\prime=j+\frac{1}{2}\hat{x}-\frac{1}{2}\hat{y}$.  Fig.~\ref{fig:cartoon} provides a cartoon of the orbitals with their phase factors, where the product of the phase factors determines $\eta_{ij}$ and $\beta_{jj^\prime}$.  This convention is not unique; other definitions of the phases will lead to the same results due to  gauge invariance.

The remaining terms in the Hamiltonian are defined as

\begin{eqnarray}
V_{dd}&=&U_{dd}\sum_i n^d_{i\uparrow}n^d_{i\downarrow}+(\epsilon_d-\mu)\sum_{i,\sigma}n^d_{i\sigma},\nonumber\\
V_{pp}&=&U_{pp}\sum_j n^p_{j\uparrow}n^p_{j\downarrow}+(\epsilon_p-\mu)\sum_{j,\sigma}n^p_{j\sigma},\nonumber\\
\end{eqnarray}
where $n^\alpha_{i \sigma}$ is the number operator of holes with spin $\sigma$ and orbital character $\alpha$ in unit cell $i$.  $U_{dd}$ and $U_{pp}$ are the strengths of the copper and oxygen on-site interactions, respectively.  Finally, the chemical potential $\mu$ controls the filling, where $\epsilon_d$ and $\epsilon_p$ are the site energies of the copper and oxygen orbitals, respectively.  With $t_{pp}$ and $\epsilon_d$ set to 0 for simplicity, the non-interacting band structure is $E_k^1=\epsilon_p-\mu$, $E_k^{2,3}=1/2(\epsilon_p-2\mu)\pm (1/2)\sqrt{\epsilon_p^2+16t_{pd}^2[\sin^2{(k_x a/2)}+\sin^2{(k_y a/2)}]}$.

Unless explicitly stated, we use a canonical parameter set for the cuprates (in units of eV): $U_{dd}=8.5$, $U_{pp}=4.1$, $t_{pd}=1.13$, $t_{pp}=0.49$, and $\epsilon_p-\epsilon_d=3.24$.  The hopping integrals were calculated by a cluster-model approach; the on-site interaction strengths were determined via the local-density method and found to be consistent with photoemission spectroscopy results.\cite{Ohta_PRB_1991,Johnston_EPL_2009,Czyzyk_PRB_1994}  Our results remain qualitatively the same with other parameter sets for the cuprates.\cite{Hybertsen_PRB_1989}  The main parameter we will vary in this study is $U_{pp}$.  Throughout, we work on square lattices, where the total number of Cu atoms (or unit cells) is denoted by $N$ (Cu$_N$O$_{2N}$). We set $a = 1$ as the unit of length and report all energies in eV unless otherwise stated. Finally, all occupancies are reported in hole language, where $\langle \mathrm{n_{tot}} \rangle = 0$ corresponds to six electrons/CuO$_2$ unit cell.  We define half filling as $\langle \mathrm{n_{tot}} \rangle = 1$, so hole doping corresponds to $\langle \mathrm{n_{tot}} \rangle > 1$ and electron doping $\langle \mathrm{n_{tot}} \rangle < 1$ in our notation.  For ease of comparison to earlier studies, however, the single-particle spectral functions are shown in electron language, where hole doping corresponds to $\langle \mathrm{n_{tot}} \rangle < 1$ and electron doping $\langle \mathrm{n_{tot}} \rangle > 1$.  The oxygen spectral weight is defined as the sum of the $2p_x$ and $2p_y$ single-particle spectra.

\subsection{DQMC algorithm}
This section provides an overview of the DQMC algorithm for the three-orbital Hubbard model.\cite{Dopf_PRB_1990,Scalettar_PRB_1991,Dopf_PRL_1992}  Observables are computed in the grand canonical ensemble

\begin{eqnarray}
\langle \hat{O} \rangle=\frac{\mathrm{tr} [\hat{O}e^{-\beta H}]}{\mathrm{tr} [e^{-\beta H}]},
\end{eqnarray}
with the imaginary-time interval $[0,\beta]$ divided into $L$ slices of width $\Delta \tau$.  Rewriting the Hamiltonian in terms of the non-interacting and interacting pieces, the exponential can be decomposed using the Trotter approximation

\begin{eqnarray}
e^{-L \Delta\tau H} \approx (e^{-\Delta\tau K_{pd}}e^{-\Delta\tau K_{pp}}e^{-\Delta\tau V_{dd}}e^{-\Delta\tau V_{pp}})^L,
\end{eqnarray}
dropping terms in the expansion of order $O(\Delta\tau^2)$ and higher.  The non-interacting terms are quadratic in the fermion operators and can be evaluated in a straightforward manner; however, the interaction terms are quartic in the fermion operators and require more care.  To transform them to quadratic form, auxiliary Hubbard-Stratonovich fields $s_{m,l}$ are introduced at each site $m$ and time slice $l$:

\begin{eqnarray}
&&e^{-\Delta\tau U_{\alpha \alpha} n_{m\uparrow}^\alpha n_{m\downarrow}^\alpha}\nonumber\\
&&=\frac{1}{2}\sum\limits_{s_{m,l}} s_{m,l} e^{\lambda_\alpha s_{m,l} (n_{m\uparrow}^\alpha - n_{m\downarrow}^\alpha) -\frac{1}{2}U_{\alpha \alpha} \Delta\tau (n_{m\uparrow}^\alpha + n_{m\downarrow}^\alpha)},
\end{eqnarray}
where $\alpha$ refers to the $d$ or $p$ orbitals and $\lambda_\alpha$ is defined by $\tanh^2{(\lambda_\alpha/2)}=\tanh{(\Delta\tau U_{\alpha \alpha} /4)}$.  Once the interaction terms have been rewritten in quadratic form, the trace over the fermion degrees of freedom can be performed and the partition function takes the form

\begin{eqnarray}
Z = \sum\limits_{s_{m,l} =\pm1} \det{M^{+}} \det{M^{-}},
\end{eqnarray}
where

\begin{eqnarray}
M^{\sigma} &=& I + B^\sigma_LB^\sigma_{L-1}...B^\sigma_1.
\end{eqnarray}

Here, 

\begin{eqnarray}
B^\pm_l = e^{-\Delta\tau k^{pd}} e^{-\Delta\tau k^{pp}} e^{v_\pm^d(l)} e^{v_\pm^p(l)},
\end{eqnarray}
with $I$ as the identity matrix and the definition

\begin{eqnarray}
v_\pm^\alpha(l)_{mm^\prime} = \delta_{mm^\prime} \Big[\pm \lambda_\alpha s_{m,l} + \Delta\tau \Big(\mu-\epsilon_\alpha - \frac{U_{\alpha \alpha}}{2} \Big) \Big].
\end{eqnarray}

The observable $\langle \hat{O} \rangle$ can be calculated via a standard Markov-chain Monte Carlo technique to sample the Hubbard-Stratonovich fields $s_{m,l}$, using a modified heat bath algorithm to accept proposed changes. The weight of each Hubbard-Stratonovich field configuration is given by $\det{M^{+}}\det{M^{-}}/Z$, but the product of determinants is not positive definite.  To ensure the probability distribution for a given configuration $\{s\}$ is positive definite, the probability is taken to be $P(s)=|\det{M^{+}}\det{M^{-}}/Z|$ and the fermion sign, $f_\mathrm{sgn}$, is included in the expression separately:

\begin{eqnarray}
\langle \hat{O} \rangle = \frac{\sum\limits_{s_{m,l}} \hat{O} f_\mathrm{sgn} P(s)}{\sum\limits_{s_{m,l}} f_\mathrm{sgn} P(s)},
\end{eqnarray}
where the quantity in the denominator is the average fermion sign.  Except in specific cases where it is protected by symmetry (such as at half filling in the single-band Hubbard model), the average fermion sign is less than 1.  As the system size increases or the temperature decreases, the average fermion sign tends towards 0, leading to an amplification of statistical fluctuations and limiting the parameter regime that can be accessed (discussed in greater detail in the following section).

In order to obtain quantities such as spectral functions $\mathrm{A_\alpha}(k,\omega)=\mathrm{G_{\alpha \alpha}}(k,\omega)$ and the density of states (DOS) to compare to experiments, the imaginary-time Green's functions are analytically continued to real frequencies.  As a note, the spectral function is given by the trace
of the Green's function matrix in either the orbital or band basis. The spectral function is positive definite and obeys a sum rule.  Here we employ  the maximum entropy method\cite{Jarrell_MEM_1996} (MaxEnt, or MEM) that uses Bayesian statistical inference to determine the most probable, or ``best," spectral density given an imaginary-time correlator.

In this study, we compute equal-time (or static) single- and two-particle correlations as well as the unequal-time spectral functions, expanding on earlier work by using a parameter set specific to the cuprates, which includes physical effects such as finite oxygen-oxygen hopping, accessing larger system sizes, and resolving orbital-dependent behavior in the spectral functions\cite{Dopf_PRB_1990,Scalettar_PRB_1991,Kuroki_PRL_1996}.  The equal-time quantities, such as the filling and spin-spin and density-density correlation functions, provide an energy-integrated perspective on how the system responds to the addition or removal of particles and to excitations.  The unequal-time quantities, such as the single-particle spectral function and density of states, complement the equal-time quantities with information on how spectra are distributed as a function of energy.  Together, they facilitate a more complete understanding of how the model behaves.

\section{Fermion sign}

\begin{figure}[b!]
	\includegraphics[width=\columnwidth]{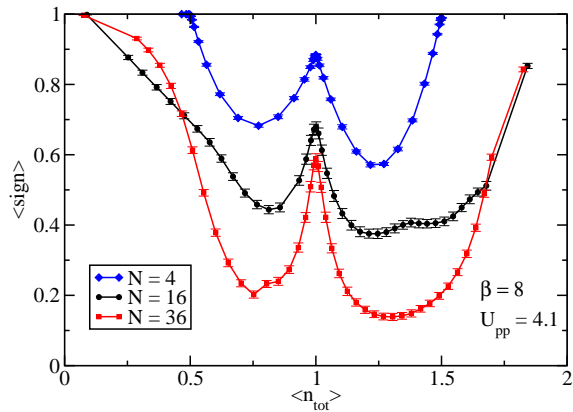}
	\caption{The average fermion sign is plotted versus filling for $N=4$, $N=16$, and $N = 36$ at $\beta=8$ eV$^{-1}$ and $U_{pp}=4.1$ eV.  In general, a larger system size results in a more severe sign problem.}
	\label{fig:sign_vs_N}
\end{figure}

The fermion sign affects the statistical error in the calculation; hence its average value determines the regions of parameter space that are accessible to the simulation.  Particle-hole symmetry is unbroken in the simple undoped single-band Hubbard model with nearest-neighbor hopping only and ensures that the average fermion sign is 1; however, this symmetry is broken in the three-orbital Hubbard model for any combination of hoppings and the fermion sign is only partially protected at $0\%$ doping.  As shown in Fig.~\ref{fig:sign_vs_N}, hole or electron doping reduces the sign to a minimum value at intermediate doping levels relevant to the cuprates.  A clear particle-hole doping asymmetry is observed, reflecting the natural asymmetry in the three-orbital Hubbard model.  At $\langle \mathrm{n_{tot}} \rangle=0$ and $2$, the symmetry between the up and down spins fully protects the fermion sign.

As demonstrated generally for quantum systems,\cite{Iazzi_PRB_2016} the average sign is proportional to $\exp{(-\beta V \Delta F)}$, where $V$ is the volume of the system and $\Delta F$ is the difference of free energy densities between the fermionic system and the corresponding bosonic system used for Monte Carlo sampling.  Statistical errors grow exponentially with increasing system size and decreasing temperature, as reflected in Fig.~\ref{fig:sign_vs_N}.  Here the volume $V=3N$, so the average sign decreases as $N$ increases.  Interestingly, the average fermion sign shows local maxima away from $0\%$ doping in the $N=4$ system, an effect which would presumably become more pronounced at lower temperatures.\cite{White_PRB_1989,Iglovikov_PRB_2015}  When $U_{pp}>0$, all the sites show correlations such that for all possible hopping paths, the order of the fermion operators would be important.  Hence, the average sign is suppressed more on the hole-doped side than the electron-doped side.  When $U_{pp}=0$, Hubbard-Stratonovich fields on oxygen are eliminated, allowing the simulation to access hole doping levels relevant to the cuprates more easily.

\begin{figure}[t!]
	\includegraphics[width=\columnwidth]{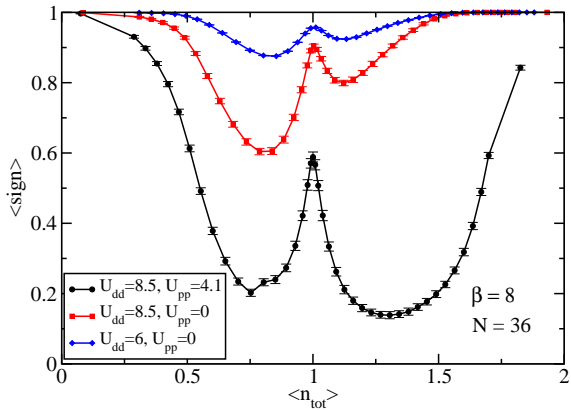}
	\caption{The average fermion sign is plotted versus filling for different values of $U_{dd}$ and $U_{pp}$ at $\beta=8$ eV$^{-1}$ and $N = 36$.  Reducing the interactions improves the sign problem and, in the case of $U_{pp}$, can even flip which side of diagram has the more severe problem.}
	\label{fig:sign_vs_U}
\end{figure}

Increasing the on-site interactions decreases the average fermion sign (Fig.~\ref{fig:sign_vs_U}), because larger values of $U_{dd}$ or $U_{pp}$ weight the Hubbard-Stratonovich field configurations, which are potentially negative, more heavily in the path integral.  Since the choice of interaction strength on the copper orbitals is guided by spectroscopy, $U_{dd}$ cannot be reduced significantly while still making a meaningful comparison with real materials.  However, because $p$-electrons are more delocalized and hence interact less strongly than $d$-electrons, there is greater leeway in selecting the oxygen on-site interaction.  It has been common practice in literature to neglect $U_{pp}$ altogether.\cite{Dopf_PRB_1990,Scalettar_PRB_1991,Dopf_PRL_1992,Dopf_PRL_1992_2}  As expected, setting $U_{pp}=0$ eliminates the Hubbard-Stratonovich fields on oxygen and hence leads to a dramatic improvement in the sign problem.  

\begin{figure}[t!]
	\includegraphics[width=\columnwidth]{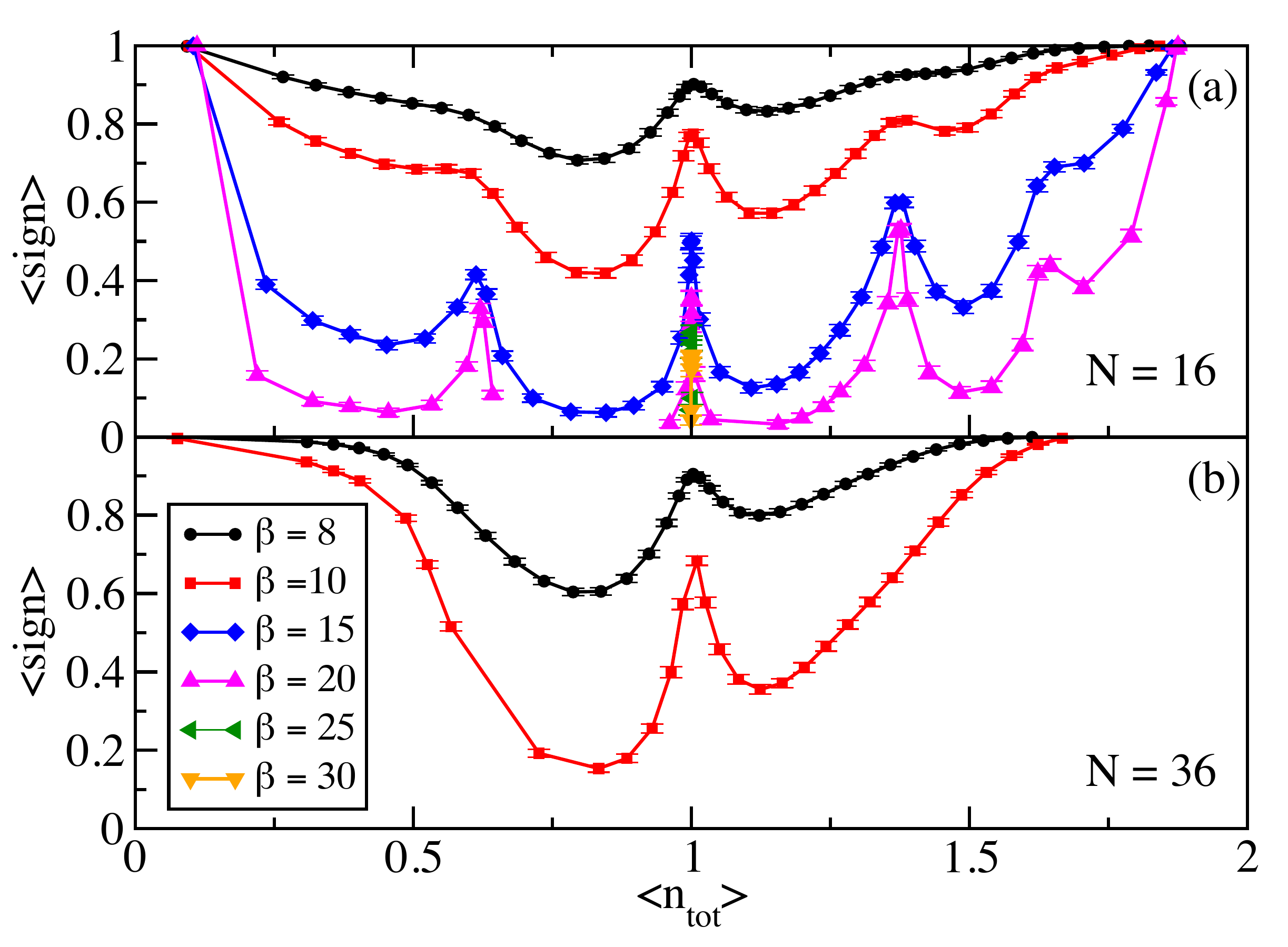}
	\caption{The average fermion sign is plotted versus filling for a range of temperatures for the (a) $N=16$ and (b) $N = 36$ systems, with $U_{pp}=0$ to access lower temperatures.  Decreasing the temperature significantly decreases the average sign but preserves the direction of the particle-hole asymmetry.  The partially protected fillings in the $N=16$ system become more pronounced with decreasing temperature.}
	\label{fig:sign_vs_B}
\end{figure}

As discussed above, the fermion sign decreases exponentially with decreasing temperature but can be enhanced by reducing the interaction strengths, or even neglecting $U_{pp}$, to access the lowest possible temperatures for a given system size.  It has been proposed that as the allowed $k$ points on small clusters fill up, local maxima occur in the average sign, which is a system geometry effect that is enhanced at lower temperatures.\cite{Iglovikov_PRB_2015}  Figure~\ref{fig:sign_vs_B}(a) shows the development of maxima in the average sign at specific doping levels on the $N=16$ cluster.  That the doping levels are determined by the system geometry is supported by the absence of local maxima at the same doping levels in the $N=36$ system [Fig.~\ref{fig:sign_vs_B}(b)] (presumably, the $N=36$ fermion sign will develop local maxima at different doping levels at lower temperatures).  This result suggests that one can tune the partially protected doping levels by changing the system geometry in order to access particular doping levels that are inaccessible on most clusters. 

Figure~\ref{fig:sign_vs_B} also demonstrates that the qualitative behavior of the electron-hole doping asymmetry is preserved, as the average sign is suppressed more by electron doping than hole doping regardless of temperature. However, the enhancement of the average sign on the hole-doped side relative to that on the electron-doped side is less pronounced as temperature decreases, making it comparatively more difficult to access doping levels relevant to the cuprates.  Together, Figs.~\ref{fig:sign_vs_U} and \ref{fig:sign_vs_B} suggest that it is possible to access lower temperatures at a specific doping level by taking advantage of the asymmetry from including or neglecting oxygen interactions.

\begin{figure}[t!]
	\includegraphics[width=\columnwidth]{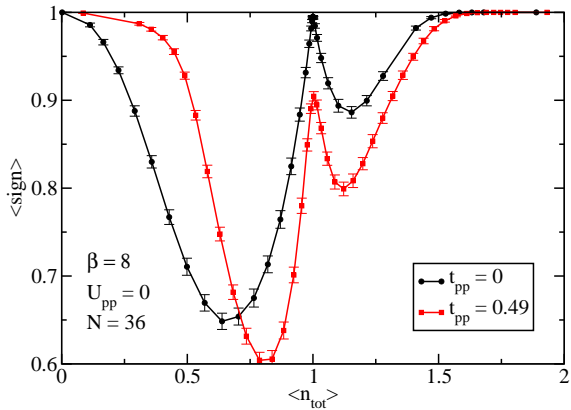}
	\caption{The average fermion sign is plotted versus filling for different values of $t_{pp}$ to demonstrate the effect of decreasing the number of hopping pathways.  Again, the particle-hole asymmetry is evident, as $t_{pp}=0$ invariably enhances the average sign for hole doping more than it does for electron doping.}
	\label{fig:sign_vs_tpp}
\end{figure}

Further improvement of the average fermion sign over a wide doping range can be achieved by setting the oxygen-oxygen hopping $t_{pp}=0$, as shown in Fig.~\ref{fig:sign_vs_tpp}.  Reducing the number of hopping pathways and hence the number of closed fermion loops will reduce the number of permutations of creation and annihilation operators that can potentially lead to negative signs.  The particle-hole doping asymmetry remains and shows the same qualitative behavior as when $t_{pp}$ is finite, with the average sign suppressed more on the electron-doped side.  In the three-orbital Hubbard model, $t_{pp}$ plays a role analogous to that of the next-nearest-neighbor hopping $t^\prime$ in the single-band Hubbard model.\cite{Eskes_PhysicaC_1989}  In both the single-band and three-orbital models, the average sign at $0\%$ doping is suppressed, and the local minimum in the average sign shifts at higher electron doping levels when $t^\prime$ or $t_{pp}$ is finite.  At low electron doping levels, the low hole density means that $t_{pp}$ has a decreased effect on the average sign.  

A systematic exploration of how different parameters affect the average fermion sign points to a few ways to improve the sign problem and simulate larger system sizes and lower temperatures while ensuring that the model is applicable to the cuprates.  The first method is to adjust the geometry of the system to shift the doping levels at which the sign has local maxima.  As shown in Fig.~\ref{fig:sign_vs_B}(a), the sign can be dramatically increased, especially at low temperatures.  In addition, a recent study of the fermion sign in the single-band model has shown that the rectangular lattice ameliorates the sign problem compared to the square lattice,\cite{Iglovikov_PRB_2015} which merits examination in the three-orbital model.  However, this approach can be computationally intensive, so the second method takes advantage of the particle-hole doping asymmetry in the average sign when $U_{pp}>0$ versus $U_{pp}=0$ to more easily access hole or electron doping levels.  A third method is to achieve an improvement in the sign for all dopings relevant to the cuprates by neglecting oxygen-oxygen hopping, as has been done in earlier studies,\cite{Dopf_PRB_1990,Scalettar_PRB_1991,Dopf_PRL_1992,Dopf_PRL_1992_2} but it risks missing important physics, such as the stability of the Zhang-Rice singlet.

\section{Energy and static correlations}
\subsection{Double occupancy and energy}
\begin{figure}[t!]
	\includegraphics[width=\columnwidth]{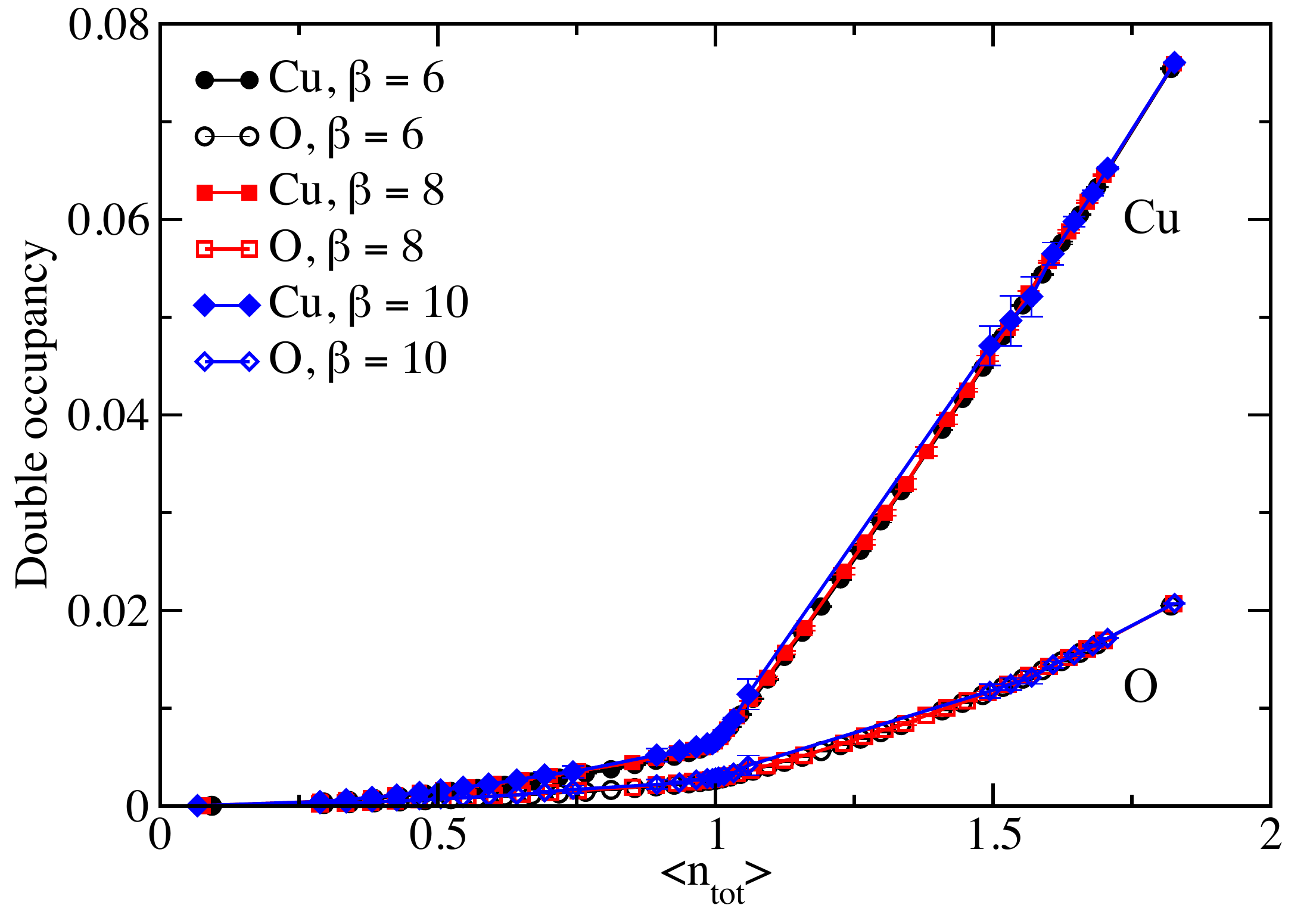}
	\caption{The double occupancy $\mathrm{D_\alpha}$ versus filling on the copper and oxygen orbitals is shown for different temperatures, with $N=36$ and $U_{pp}=4.1$ eV.  It exhibits no significant system size or temperature dependence.}
	\label{fig:double_occ}
\end{figure}

In this section, the doping, system size, and temperature dependences of the equal-time double occupancy and energy of the three-orbital Hubbard model are explored systematically.  Figure~\ref{fig:double_occ} shows the orbitally-resolved double occupancies, $\mathrm{D_\alpha} = \sum_i \langle n^\alpha_{i\uparrow} n^\alpha_{i\downarrow} \rangle$ with $\alpha$ as the orbital index, versus filling for different temperatures on the $N=36$ system.  The undoped system has on average $\langle \mathrm{n_{Cu}} \rangle \sim 0.7-0.8$ and $\langle \mathrm{n_{O_{x,y}}} \rangle \sim 0.2-0.3$ [see Fig.~\ref{fig:filling}(b)].  Hence, any doped holes that reside on copper will add to the double occupancy, while doped holes that go into an oxygen orbital will in general not encounter a pre-existing hole.  As a result, the double occupancy increases much more rapidly on the copper orbitals than on the oxygen orbitals.  On the electron-doped side, the orbitally-resolved double occupancies change less rapidly than on the hole-doped side of the phase diagram because there were fewer to start with, as in general each copper orbital has 0 or 1 holes.  Lowering the temperature from $\beta=6$ to $\beta=10$ eV$^{-1}$ or changing the system size from $N=36$ to $N=16$ barely affects the double occupancy.  Hence the potential energy, which is a sum of the double occupancies on copper and oxygen orbitals weighted by the on-site interactions, is essentially independent of temperature and system size.

\begin{figure}[t!]
	\includegraphics[width=\columnwidth]{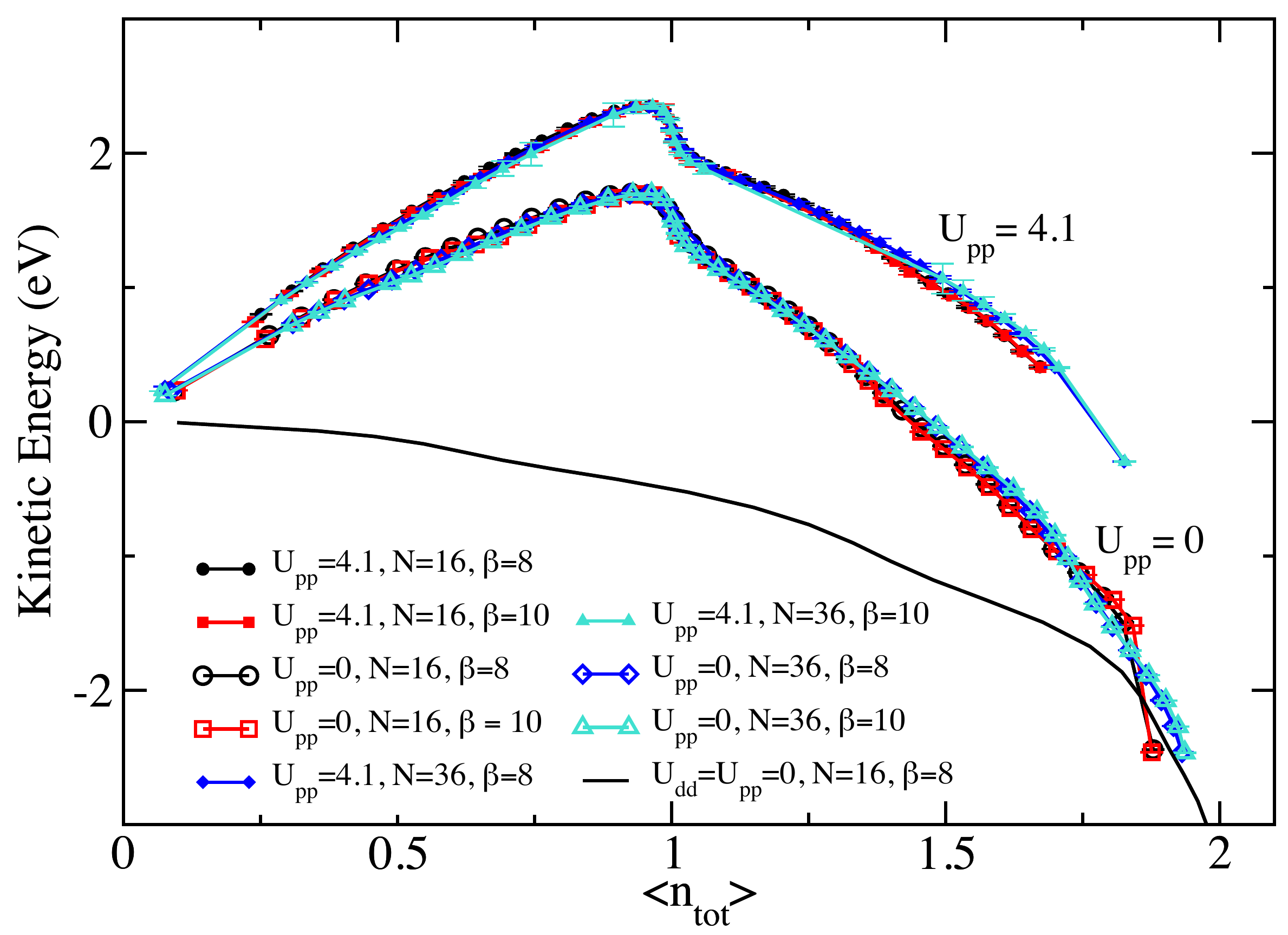}
	\caption{Kinetic energy of the holes versus filling for different temperatures in $N=16$ and $N=36$ systems, with $U_{pp}=4.1$ and $0$ eV.  The non-interacting kinetic energy (solid black line) is shown for comparison.}
	\label{fig:KE}
\end{figure}

The single-particle kinetic energy $K= \langle K_{pd} \rangle + \langle K_{pp} \rangle$ is governed by two competing trends with hole doping in the correlated system.  It is increased by having more holes available to hop, and it is decreased by double occupancies that block hopping pathways.  Figure~\ref{fig:KE} shows that the kinetic energy steadily increases from $\langle \mathrm{n_{tot}} \rangle=0$ to $\langle \mathrm{n_{tot}} \rangle=1$ as the addition of holes increases hopping.  It reaches a maximum at $\langle \mathrm{n_{tot}} \rangle=1$ before decreasing again due to increasing double occupancy (Fig.~\ref{fig:double_occ}).  The effect of double occupancies can be seen especially clearly from the low kinetic energy near $\langle \mathrm{n_{tot}} \rangle=2$, where the lowest band is filled.  The abrupt decrease near $0\%$ doping corresponds to the Mott gap in the filling (see next section).   Setting $U_{pp}=0$ removes the penalty on double occupancy for oxygen, so there is reduced impetus for holes to hop off oxygen orbitals and hence a lower kinetic energy.  Like the potential energy, the kinetic energy does not have a strong dependence on system size or temperature.  For comparison, the kinetic energy in the non-interacting system is included, showing that there is no Mott gap and hence no abrupt decrease in the energy near $0\%$ doping when $U_{dd}=U_{pp}=0$.  As in the single-band Hubbard model, the Mott physics away from $0\%$ doping is not very prominent from the single-particle perspective.

\subsection{Filling}

\begin{figure}[t!]
	\includegraphics[width=\columnwidth]{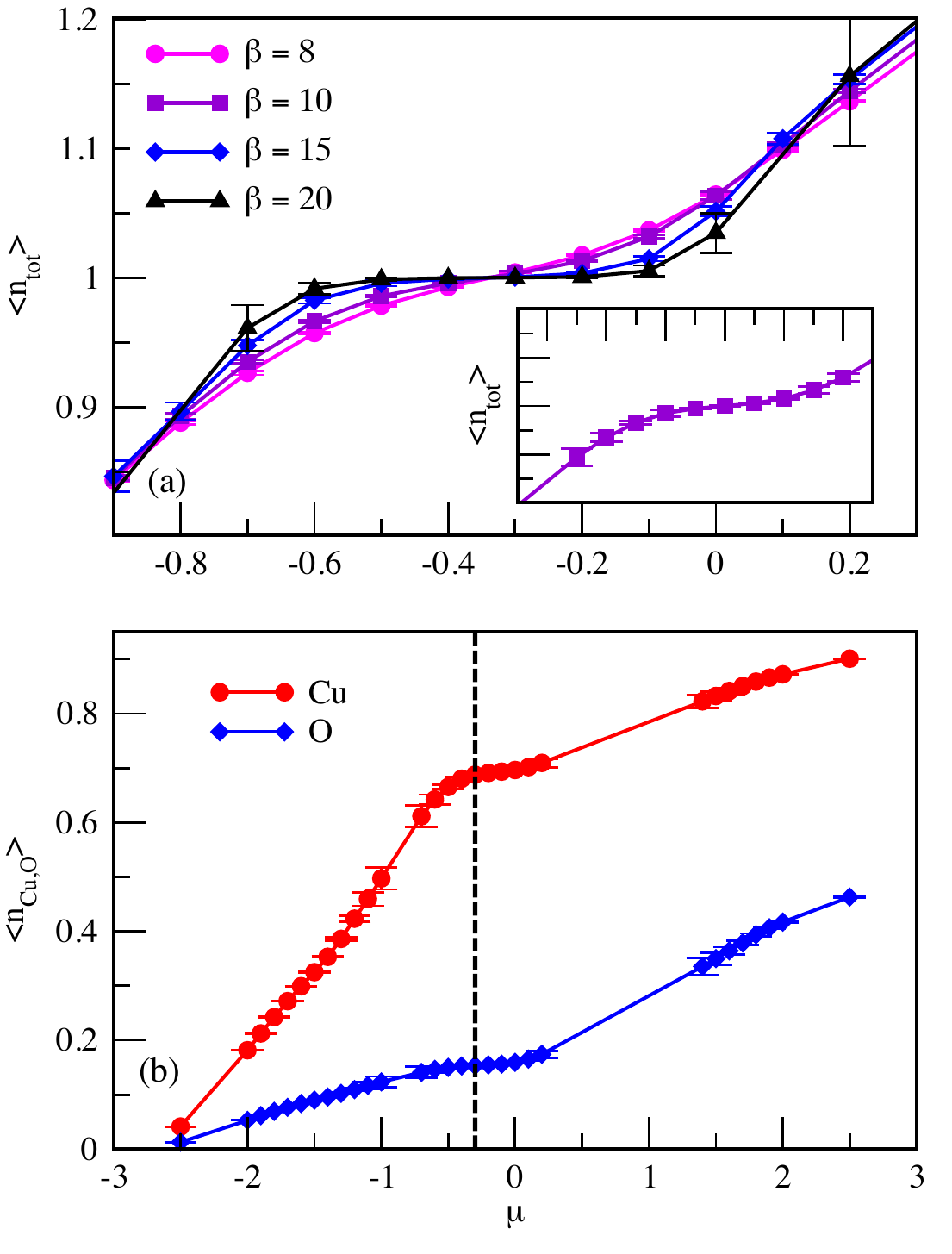}
	\caption{(a) The total filling curve shows a gap opening with decreasing temperature, with $U_{pp}=0$ eV and $N=16$ to access the lowest possible temperatures.  The inset shows the total filling for $N=36$, $U_{pp}=4.1$ eV, and $\beta=10$ eV$^{-1}$ and has the same axes.  (b) The orbitally resolved fillings are shown for the same parameters as the inset in (a) and demonstrate that doped holes preferentially reside on oxygen atoms. As we are using the larger system size, the fermion sign is too small at certain chemical potentials to determine the filling; however, the solid lines indicate the trend that is consistent with results in smaller systems.  A dashed line indicates the chemical potential corresponding to the undoped system.}
	\label{fig:filling}
\end{figure}

As mentioned above, equal-time quantities provide information on energy-integrated, static correlations.  In particular, the filling, which is a single-particle quantity, demonstrates how the system responds to the addition or removal of particles.  The total filling $\langle \mathrm{n_{tot}} \rangle$ can be studied as a function of the chemical potential $\mu$  to explore the opening of a gap at low temperatures.  We set $U_{pp}=0$ and $N=16$ to access temperatures as low as $\beta=20$ eV$^{-1}$ in a window around $0\%$ doping [Fig.~\ref{fig:filling}(a)] to show that a distinct plateau, corresponding to the Mott gap, develops as temperature decreases.  To confirm that neglecting oxygen on-site interactions and using a smaller system size do not affect the conclusions, the total filling with $U_{pp}=4.1$ eV, $N=36$, and $\beta=10$ eV$^{-1}$ is shown in the inset.  The gap appears to open at higher temperatures in the larger system, but overall the qualitative behavior is similar, indicating that oxygen on-site interactions do not play a significant role close to the Fermi level ($E_F$). 

Figure~\ref{fig:filling}(b) shows the fillings on the copper and oxygen $p_x$ orbitals (which have the same behavior as the oxygen $p_y$ orbitals due to $x-y$ symmetry in the model) for the $U_{pp}=4.1$ eV, $N=36$, and $\beta=10$ eV$^{-1}$ system.  The orbitally-resolved fillings exhibit distinct particle-hole asymmetry in their slopes.  As known in the cuprates, doped holes preferentially reside on oxygen, which is reflected in the $\langle \mathrm{n_O} \rangle$ doping trend.  Doped electrons, on the other hand, generally reside on the copper orbitals; hence $\langle \mathrm{n_{Cu}} \rangle$ has a higher slope on the electron-doped side.  This behavior is consistent for different system sizes and temperatures, indicating that the occupation is independent of these simulation details.

\subsection{Spin-spin correlation function}
\begin{figure}[t!]
	\includegraphics[width=\columnwidth]{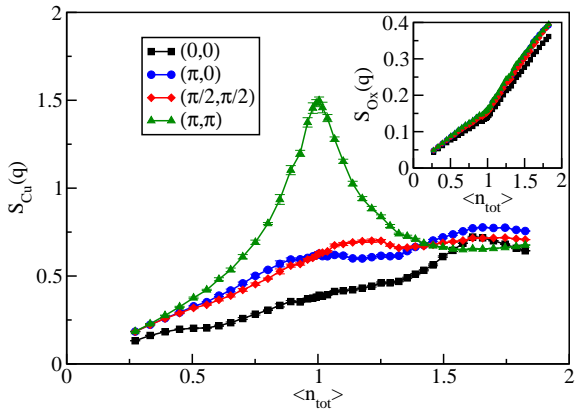}
	\caption{The spin-spin correlation function is plotted versus filling for four different possible ordering vectors on the copper and oxygen (inset) orbitals.  On copper, $(\pi,\pi)$ antiferromagnetism dominates in the undoped system, decreasing with doping.  The oxygen orbitals do not show signs of any particular spin order.  Parameters used here are $N=16$, $U_{pp}=4.1$ eV and $\beta=10$ eV$^{-1}$.}
	\label{fig:Sq_vs_fill}
\end{figure}

To understand how the system responds to excitations or perturbations, it is necessary to examine multi-particle quantities.  In addition, single- and multi-particle quantities have been seen in the single-band Hubbard model to exhibit different renormalizations with doping,\cite{Kung_PRB_2015} suggesting that it may be important to study two-particle quantities such as the spin-spin and density-density correlations to understand the behavior of the three-orbital system.  The orbitally-resolved equal-time spin-spin correlation function is defined as

\begin{eqnarray}
S_\alpha (\mathbf{q})&=&\sum_l e^{i\mathbf{q}\cdot \mathbf{l}} S_\alpha (l_x,l_y),
\end{eqnarray}
where
\begin{eqnarray}
S_\alpha (l_x,l_y)&=&\frac{1}{N}\sum_i \langle (n_{i \uparrow}^\alpha-n_{i\downarrow}^\alpha)(n_{i+l \uparrow}^\alpha-n_{i+l \downarrow}^\alpha) \rangle.
\end{eqnarray}
The spin-spin correlation function on copper orbitals shows a pronounced tendency towards N\'{e}el antiferromagnetic ordering, especially near $0\%$ doping.  As shown in Fig.~\ref{fig:Sq_vs_fill}, the ordering vector $\mathbf{q}=(\pi,\pi)$ dominates for a wide doping range (approximately $60\%$ electron doping to $40\%$ hole doping), although the antiferromagnetic tendency is destroyed rapidly with increasing doping in agreement with experiment.  The particle-hole doping asymmetry also agrees with experiments showing that AFM is more robust on the electron-doped side of the phase diagram.\cite{Armitage_RMP_2010}  For these geometries, there is no obvious sign of oxygen spin lattice symmetry breaking around $12.5\%$ hole doping (inset of Fig.~\ref{fig:Sq_vs_fill}).
\begin{figure}[t!]
	\includegraphics[width=\columnwidth]{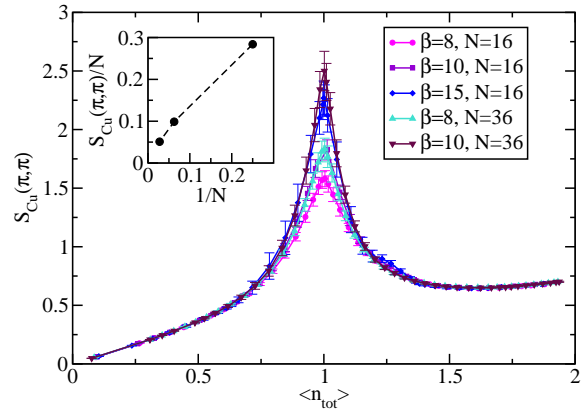}
	\caption{The copper spin-spin correlation function shows that $\mathbf{q}=(\pi,\pi)$ order is strengthened by increasing system size and decreasing temperature ($U_{pp}=0$).  The inset shows the finite size scaling at $0\%$ doping and $\beta=8$ eV$^{-1}$.}
	\label{fig:Spipi}
\end{figure}

Figure~\ref{fig:Spipi} focuses on $(\pi,\pi)$ AFM, which has been well characterized by experiments, and demonstrates that $\mathrm{S_{Cu}}(\pi,\pi)$ peaks more strongly as the system size increases or as the temperature decreases.  This behavior suggests that DQMC simulations can identify trends at higher temperatures that correspond to low-temperature ordered phases without necessarily accessing the thermodynamic limit or the low temperatures comparable to those in experiments.  Finite size scaling demonstrates that $\mathrm{S_{Cu}}(\pi,\pi)/N$ tends to 0 as $1/N$ decreases (inset of Fig.~\ref{fig:Spipi}),\cite{Dopf_PRB_1990} indicating that there is no true long-range antiferromagnetic order at finite temperatures, as expected in two dimensions by the Mermin-Wagner theorem.  A comparison of Fig.~\ref{fig:Sq_vs_fill}, where $U_{pp}$ is finite, and Fig.~\ref{fig:Spipi}, where $U_{pp}=0$, indicates that there is no qualitative difference in the behavior of the copper spin-spin correlation function and that the effects of oxygen interactions are negligible.  

\begin{figure}[t!]
	\includegraphics[width=\columnwidth]{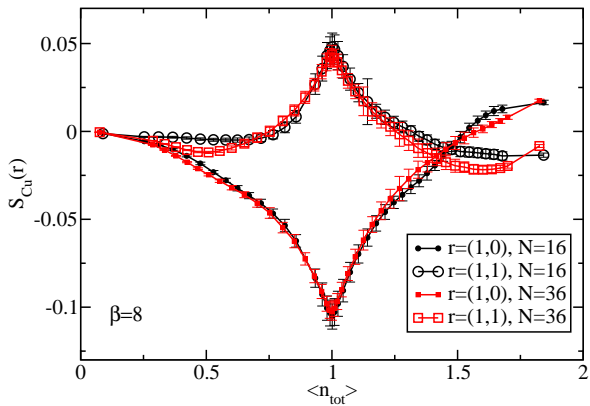}
	\caption{Nearest [$\mathrm{S_{Cu}}(1,0)$] and next-nearest [$\mathrm{S_{Cu}}(1,1)$] neighbor Cu-Cu spin-spin correlation function versus filling, with $U_{pp}=4.1$ eV and $\beta=8$ eV$^{-1}$, showing the crossover from AFM to short-ranged FM correlations for $N=16$ and $36$.}
	\label{fig:Sr}
\end{figure}

Examining the spin-spin correlation function in real space provides insights into how the system crosses over from short-range antiferromagnetic to ferromagnetic correlations at high ($>40\%$) hole doping levels, an effect which has been seen in the single-band Hubbard model.\cite{Jia_NatComm_2014}  The simulations average the correlation functions over the system, so here we examine the average spin correlations.  Figure~\ref{fig:Sr} shows the nearest [$\mathrm{S_{Cu}}(1,0)$] and next-nearest neighbor [$\mathrm{S_{Cu}}(1,1)$] Cu-Cu spin correlations in the $N=16$ and $N=36$ systems.  At $0\%$ doping, $\mathrm{S_{Cu}}(1,0)$ is strongly anti-aligned with the reference spin at $(0,0)$ and $\mathrm{S_{Cu}}(1,1)$ is strongly aligned with the reference spin, which is consistent with $(\pi,\pi)$ antiferromagnetic order.  As the system is hole doped, the magnitudes of the spin correlations decrease as $(\pi,\pi)$ AFM is destroyed.  In an intermediate window, there is no particular spin order, but at high hole doping levels, $\mathrm{S_{Cu}}(1,0)$ becomes positive, indicating that the system is developing short-range ferromagnetic correlations.  Increasing the system size does not significantly impact the qualitative behavior of the correlations or even doping level of the crossover.  

The crossover from short-range antiferromagnetic to ferromagnetic correlations is not affected by temperature on a qualitative or quantitative level.  As expected, decreasing the temperature enhances the magnitudes of $\mathrm{S_{Cu}}(1,0)$ and $\mathrm{S_{Cu}}(1,1)$ near $0\%$ doping, confirming that antiferromagnetic order is strengthened.  However, even when the temperature is lowered from $\beta=8$ to $\beta=20$ eV$^{-1}$ in the $N=16$ and $U_{pp}=0$ system, there is no significant change in the doping level at which the system begins to exhibit short-ranged ferromagnetic correlations.  Hence the conclusions drawn from higher-temperature simulations appear to be robust.

Because the magnetic properties of the cuprates have been studied so intensively, the spin correlations provide an effective test of how closely DQMC simulations of the Hubbard model capture the behavior of the materials.  Neglecting $U_{pp}$ has no qualitative effect on the spin-spin correlation function and demonstrates that, at least in the equal-time quantities, oxygen does not play a significant role in the spin physics.  The results agree with observations that the cuprates exhibit N\'{e}el order near $0\%$ doping, which is more robust on the electron-doped side of the phase diagram,\cite{Armitage_RMP_2010} showing that the simulations are capable of identifying trends that correspond to low-temperature states.

\subsection{Density-density correlation function}
The question of whether the three-orbital Hubbard model shows charge order on the copper or oxygen orbitals can be addressed using the equal-time density-density correlation function

\begin{figure}[b!]
	\includegraphics[width=\columnwidth]{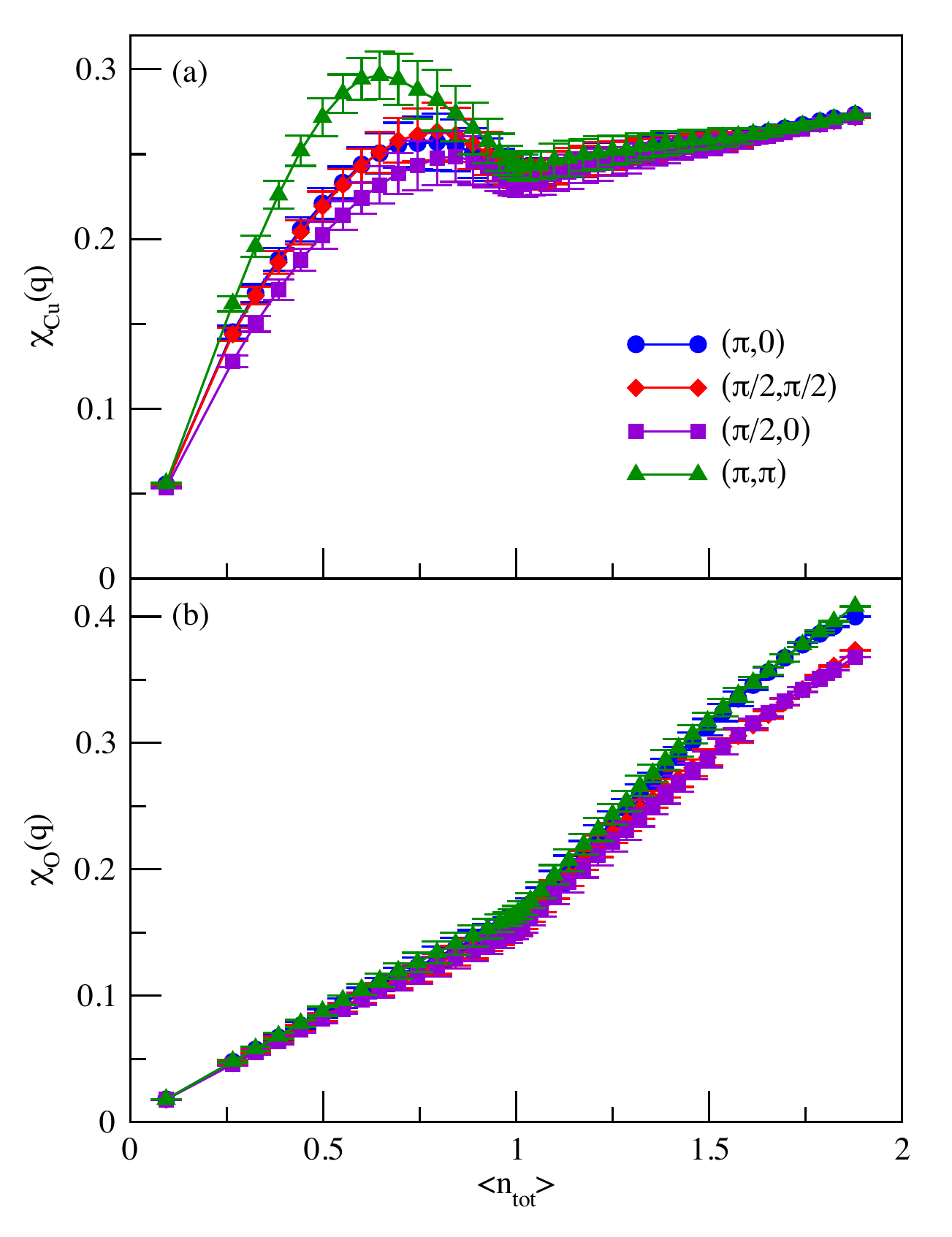}
	\caption{Density-density correlation function versus filling on the (a) copper and (b) oxygen orbitals for different ordering vectors with $N=16$ and $U_{pp}=0$ (qualitatively similar results are obtained for $U_{pp}=4.1$ eV).  The system shows a slight tendency to $(\pi,\pi)$ charge ordering on the electron-doped side on copper.}
	\label{fig:Nq}
\end{figure}
\begin{figure}[t!]
	\includegraphics[width=\columnwidth]{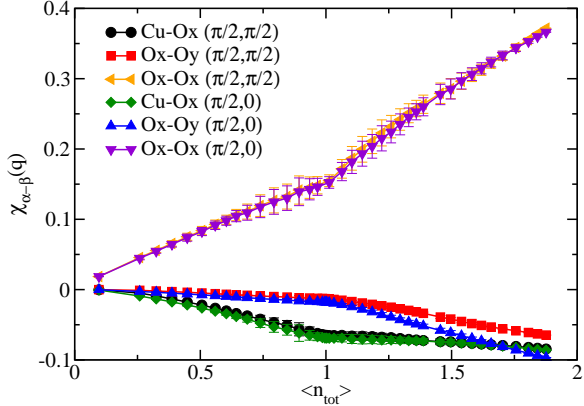}
	\caption{Density-density correlations between the copper and oxygen orbitals, and between the oxygen orbitals for different ordering vectors in a $N=16$, $U_{pp}=0$ system (qualitatively similar results are obtained for $U_{pp}=4.1$ eV).}
	\label{fig:Nq_crossterms}
\end{figure}

\begin{eqnarray}
\chi_\alpha (\mathbf{q})&=&\sum_l e^{i\mathbf{q}\cdot \mathbf{l}} \chi_\alpha (l_x,l_y),
\end{eqnarray}
where
\begin{eqnarray}
\chi_\alpha (l_x,l_y)&=&\frac{1}{N}\sum_i \langle (n_{i \uparrow}^\alpha+n_{i \downarrow}^\alpha)(n_{i+l \uparrow}^\alpha+n_{i+l \downarrow}^\alpha) \rangle.
\end{eqnarray}
The orbitally-resolved density-density correlation functions for $N=16$ and $U_{pp}=0$ are shown in Fig.~\ref{fig:Nq}.  (Qualitatively similar results are obtained for $U_{pp}=4.1$ eV.)  Despite experimental evidence for charge and spin stripes around $12.5\%$ hole doping,\cite{Tranquada_Nature_1995,Zaanen_PRB_1989,Emery_PNAS_1999,Emery_PhysicaC_1993,Machida_PhysicaC_1989,Machida_JPSJ_1990} the ordering vector $\mathbf{q}=(\pi/2,0)$ does not dominate on the copper orbitals, as would be expected.  This may be because the system geometries under consideration do not support the full stripe order pattern, which would require a width of at least 8 unit cells.  On the electron-doped side, there is a slight tendency on the copper orbitals to enhanced charge fluctuations near $(\pi,\pi)$ at $\sim30\%$ electron doping, although it is an unrealistic doping in the cuprates.  For completeness, the other non-zero elements of the density-density correlation function, $\chi_\mathrm{Cu-O_x} (\mathbf{q})$, $\chi_\mathrm{O_x-O_x} (\mathbf{q})$, and $\chi_\mathrm{O_x-O_y} (\mathbf{q})$, are shown in Fig.~\ref{fig:Nq_crossterms}.    Evidently, the oxygen orbitals show no sign of anomalous behavior or indeed of any particular charge order, either relative to other oxygen orbitals or to copper orbitals.

\section{Spectral functions}

\begin{figure}[t!]
	\includegraphics{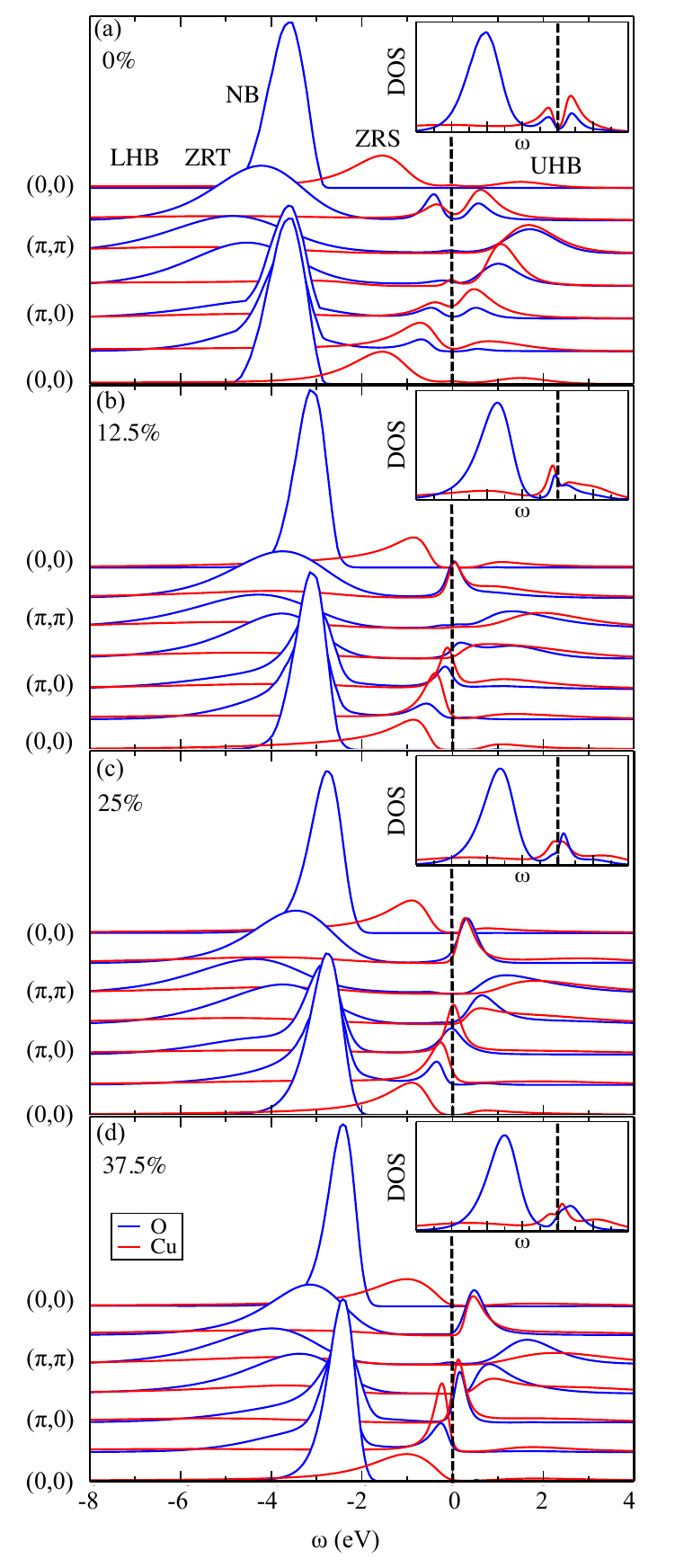}
	\caption{The orbitally-resolved spectral functions at (a) $0\%$, (b) $12.5\%$, (c) $25\%$, and (d) $37.5\%$ hole doping illustrate the doping evolution of the lower Hubbard band (LHB), Zhang-Rice triplet (ZRT) band, non-bonding (NB) band, Zhang-Rice singlet (ZRS) band, and upper Hubbard band (UHB) in the $N=16$ system, where $\beta=8$ eV$^{-1}$, $t_{pp}=0.49$ eV, and $U_{pp}=4.1$ eV.  The insets show the density of states for each doping level, with the same frequency axis as the spectral functions.}
	\label{fig:Akw}
\end{figure}

Unequal-time quantities, such as the spectral function, provide dynamical information about the behavior of a system.  The orbitally-resolved single-particle spectral function, $\mathrm{A_\alpha}(\mathbf{k},\omega)$, can be computed and compared to photoemission spectroscopy (PES) results to evaluate how well the simulations describe the cuprates.  PES has shown that the cuprates have little spectral weight near the Fermi level ($\omega=0$), with most of the weight contained in a large peak centered around $\omega=4$ eV below $E_F$ that includes contributions from the apical oxgen orbitals, non-bonding oxygen and other planar orbitals, and the Zhang-Rice triplet (ZRT) band.\cite{Fujimori_PRB_1987}  Figure~\ref{fig:Akw}(a) shows the copper (red) and oxygen (blue) spectral functions at momenta along high-symmetry cuts in the first Brillouin zone in an undoped $N=16$ system.  

\begin{figure*}[t!]
	\includegraphics[width=\textwidth]{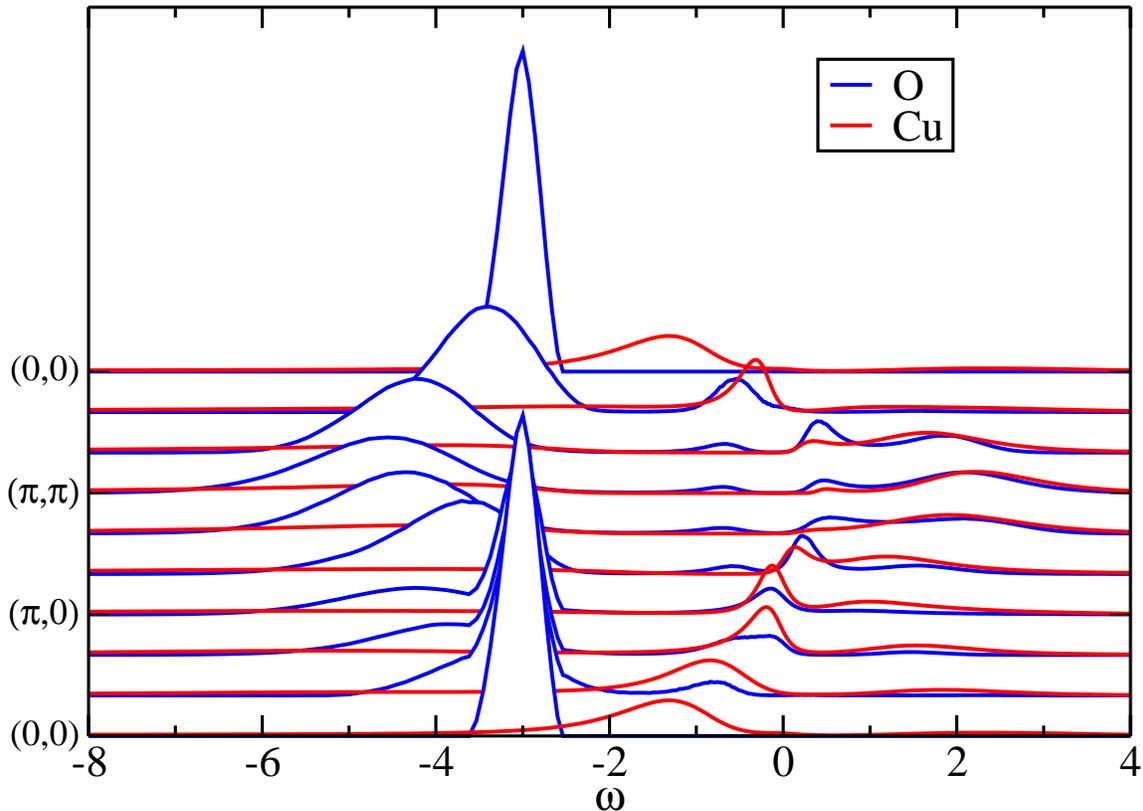}
	\caption{The orbitally-resolved spectral functions at $12.5\%$ hole doping are shown for high-symmetry cuts in the first Brillouin zone in the $N=36$ system, where $\beta=8$ eV$^{-1}$, $t_{pp}=0.49$ eV, and $U_{pp}=4.1$ eV.}
	\label{fig:Akw_Ncu_6}
\end{figure*}

The spectral functions capture the prominent PES features, with a large non-bonding (NB) band (from the non-bonding phase of the planar oxygen orbitals) with mostly oxygen orbital content around 3-4 eV below the Fermi level.  The size of the peak in PES is larger than that in our calculation because the three-orbital Hubbard model excludes not only apical and other non-bonding oxygen orbitals, but also out-of-plane oxygen $p_z$ orbitals and other $d$ orbitals  that would contribute to the non-bonding peak and low-energy structure.  Due to the high simulation temperature, the weights of the lower Hubbard band (LHB) and ZRT band are spread out into the long tails of the peaks below $\omega=4$ eV.  Similarly, the upper Hubbard band (UHB) is smeared into broad peaks and long tails above the Fermi level.  The ZRS band is located just below $E_F$.  As expected, the system has an indirect charge transfer gap between $(\pi/2,\pi/2)$ and $(\pi,0)$.\cite{Imada_RMP_1998,Zaanen_PRL_1985}  The DOS, shown in the inset, confirms that most of the spectral weight in the system resides in the oxygen NB band.  There is a clear gap at the Fermi level, but the spectral weight is non-zero due to the elevated temperatures.

Figures~\ref{fig:Akw}(b)-(d) show the spectral functions from $12.5\%$ to $37.5\%$ hole doping (optimally doped to overdoped), enabling us to identify the evolution of orbital character in different bands.  An examination of the ZRS band shows the evolution of the ZRS upon hole doping.  At $0\%$ doping, there is greater oxygen than copper weight in the peak at $(\pi/2,\pi/2)$.  Hole doping causes the copper and oxygen character to become roughly equivalent, as expected for a singlet configuration,\cite{Unger_PRB_1993,Pothuizen_PRL_1997,Yin_PRL_2008} and the ZRS appears to persist even at $37.5\%$ hole doping, lending support to the perspective that the ZRS picture is still valid at high doping levels.\cite{Chen_PRB_2013,YJChen_PRB_2013,Brookes_PRL_2015}  There is never any oxygen orbital content in the ZRS band at the $\Gamma$ point,\cite{Unger_PRB_1993,Pothuizen_PRL_1997,Yin_PRL_2008} as known from experiments with low photon energies that are more sensitive to oxygen, whereas there is a shift towards greater copper-like spectral character at higher energies, which in experiments are more sensitive to copper.\cite{Inosov_PRL_2007} These comparisons provide additional evidence for the effectiveness of the three-orbital Hubbard model in capturing cuprate physics.

\begin{figure*}[t!]
	\includegraphics[width=\textwidth]{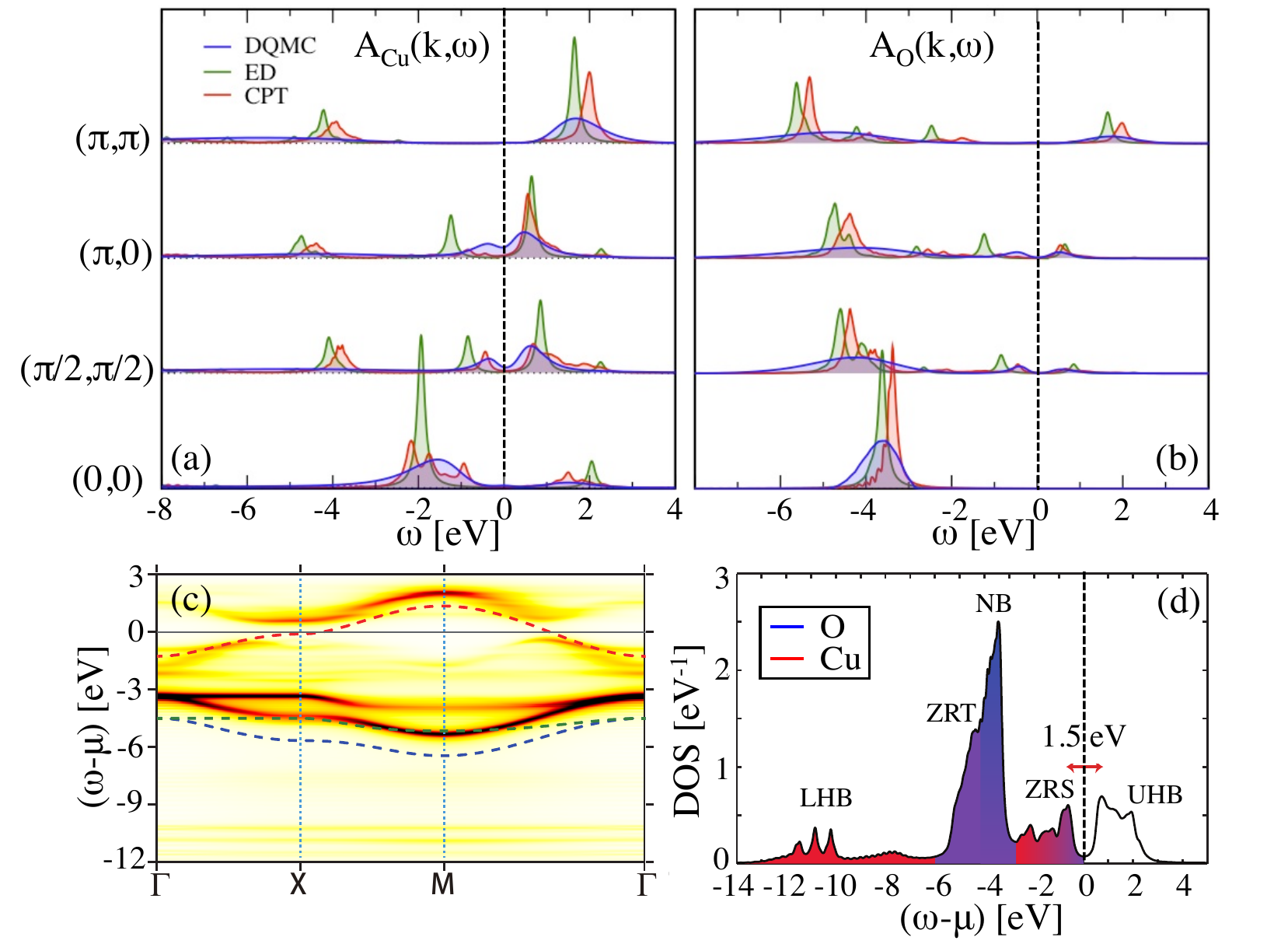}
	\caption{The (a) copper and (b) oxygen spectral functions are calculated in the undoped system using DQMC, ED, and CPT.  Despite the different system sizes and temperatures, the peak positions show reasonable agreement.  With fine momentum resolution, CPT resolves the (c) band dispersion and (d) DOS in detail.  In (c), the non-interacting bands are overlaid as dashed lines on the CPT band structure.  In (d), the shading in the DOS indicates orbital content (red for copper, blue for oxygen) in the filled bands. For example, the color gradient in the ZRS band indicates the increasing copper content away from $E_F$.}
	\label{fig:DQMC-ED-CPT}
\end{figure*}

Although accessible system sizes are limited by the fermion sign problem, we can compute the spectral functions of the $N=36$ system and access different momenta to complement the $N=16$ system.  Figure~\ref{fig:Akw_Ncu_6} focuses on $12.5\%$ hole doping, which can be compared to Fig.~\ref{fig:Akw}(b).  While shared momenta [such as $(\pi,\pi)$ and $(\pi,0)$] provide a way to gauge the impact of system size on the spectral functions, the additional points [such as $(\pi/3,\pi/3)$ and $(2\pi/3,2\pi/3)$] enable a finer momentum resolution.  Focusing again on the ZRS band, Figs.~\ref{fig:Akw}(b) and ~\ref{fig:Akw_Ncu_6} show that the oxygen content has a strong momentum dependence, which is due to a form factor that arises from the finite spatial extent of the ZRS.\cite{Katagiri_PRB_2011}  

The orbitally-resolved spectral functions thus enable a direct comparison to PES experiments to assess how well the three-orbital Hubbard model describes cuprate physics.  Clearly, including oxygen is crucial to describing the distribution of spectral weight away from the Fermi level; in fact, including the oxygen $2p_{x,y}$ orbitals is not sufficient to account for all the weight observed by PES in the NB band.\cite{Fujimori_PRB_1987}  Quantitative considerations aside, the three-orbital Hubbard model does exhibit the appropriate qualitative behavior and provides a more accurate picture of the cuprates than the single-band Hubbard model, as the latter treats the single-band LHB as the ZRS band and ignores the NB band, ZRT band, and proper LHB.\cite{Wang_PRB_2015}  The three-orbital simulations not only identify different bands, but go a step further to elucidate the evolving orbital character as the system is hole doped across the range accessible to the cuprates.

\section{Comparison to ED and CPT}
Computing quantities such as the equal-time correlation functions and single-particle spectral function enables a direct comparison of DQMC with the complementary numerical techniques of ED and CPT.  Both ED and the small-cluster calculation step of CPT simulations are performed on Cu$_8$O$_{16}$ clusters, while the DQMC calculations are performed on a Cu$_{16}$O$_{32}$ ($N=16$) system.  ED and CPT both work in a fixed particle number sector (in the canonical ensemble) at zero-temperature, while the DQMC simulation is carried out in the grand canonical ensemble at $\beta=8$ eV$^{-1}$.  All simulations are performed with the same parameter set as that given in Section II.

The orbitally-resolved spectral functions provide a consistency check between the three methods in the undoped system.  Figures~\ref{fig:DQMC-ED-CPT}(a)-(b) show that on a qualitative level, the DQMC, ED, and CPT spectra line up well at momenta on high-symmetry cuts in the first Brillouin zone.  (All three sets of spectra have been normalized to the same sum rule for ease of comparison.)  Despite thermal broadening from high temperature, DQMC peak positions are in many cases almost identical to those of CPT, especially at $(\pi/2,\pi/2)$.  Figure~\ref{fig:DQMC-ED-CPT}(c) illustrates the fine momentum resolution capability of CPT and together with the DOS in Fig.~\ref{fig:DQMC-ED-CPT}(d) enables clear identification of the different bands and their dispersions.  Unlike the DQMC simulation, in which the broad peaks and long tails make it impossible to clearly distinguish all of the bands [compare to Fig.~\ref{fig:Akw}(a)], CPT resolves the LHB around $\omega \sim 10$ eV and the ZRT band as a shoulder at $\omega \sim 5$ eV below the Fermi level.  As expected, the NB band with predominantly oxygen character is located at $\omega \sim 4$ eV below $E_F$ and the ZRS band and UHB are just below and above the Fermi energy, respectively.  

To distinguish behavior due to single-electron physics from band renormalizations due to strong correlations, the non-interacting bands are overlaid on the CPT data as dashed lines in Fig.~\ref{fig:DQMC-ED-CPT}(c).  Away from the Fermi level, the interactions transfer weight into the LHB and shift band energies without strongly affecting the qualitative behavior of the dispersion, as seen near the top of the UHB around $(\pi,\pi)$ (the M-point).  The NB band dispersion is controlled by the oxygen-oxygen hopping and hence is also not changed significantly apart from an energy shift due to $U_{pp}$.  Near the Fermi level, similar to the single-band model, interactions open a gap with precursors just above and below $E_F$ that will develop with hole doping into a ZRS band (analogous to the quasiparticle band in the single-band model) crossing the Fermi level.  In addition, as in the single-band model, interactions lead to a ``waterfall" feature that gives rise upon hole doping to the high-energy anomaly (HEA) seen in PES that ranges from approximately 400 meV to 1 eV below $E_F$ near $(\pi/4,\pi/4)$.\cite{Moritz_NJP_2009} 

The direct (optical) \cite{Tokura_PRB_1990,Cooper_PRB_1990,Uchida_PRB_1991,Falck_PRL_1992} gap in the cuprate parent compounds is known from experiments to be $1.5-2$ eV, and the indirect gap is somewhat smaller, providing a quantitative way to compare the three numerical techniques.  In ED, the direct optical gap is $\sim1.7$ eV, in agreement with experiments.  As expected, the first electron removal state occurs at $(\pi/2,\pi/2)$ and the first electron addition state at $(\pi,0)$, with a slightly smaller indirect gap of $\sim 1.5$ eV.  In  CPT, the indirect gap is determined from the peak-to-peak distance in the DOS [Fig.~\ref{fig:DQMC-ED-CPT}(d)] and is slightly smaller than the ED indirect gap.  This difference most likely arises from finite-size effects in ED, as CPT provides a better estimate of the infinite-lattice limit.  The direct and indirect gaps from the DQMC spectral functions, determined from the peak-to-peak distances in the spectral functions, appear to be smaller ($\sim1$ eV).  This discrepancy merits further investigation to determine whether it can be explained simply as a thermal or finite-size effect.

\begin{figure}[t!]
	\includegraphics[width=\columnwidth]{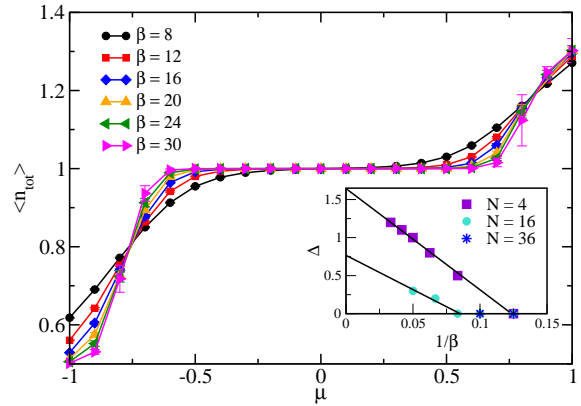}
	\caption{The filling versus chemical potential curves are shown on an $N=4$ cluster in order to access low temperatures to highlight the opening of a charge transfer gap.  The inset shows an extrapolation of the gap size $\Delta$ to zero-temperature.}
	\label{fig:fill_Ncu_2}
\end{figure}

One possible cause of the smaller DQMC gaps is peak broadening from high temperature, but another contributing factor might be the difference in system size.  Because the indirect gap requires a larger cluster to capture both the appropriate momentum points and the behavior near them, ED and CPT may overestimate the gap size.  To examine the effects of temperature and system size, we set $N=4$ in the DQMC computation to access low temperatures for a better comparison to the ED and CPT calculations, and simulate temperatures ranging from $\beta=8$ to $\beta=30$ eV$^{-1}$ (Fig.~\ref{fig:fill_Ncu_2}).  We use the width of the plateau in the filling as a proxy for the charge transfer gap because the  plateau width captures the correct chemical potential scale and the exact gap, whereas analytic continuation may not properly describe the spectral tails near the gap.  The zero-temperature extrapolated gap, which in the $N=4$ system must be the direct gap at $(\pi,0)$, is $\sim 1.6$ eV, in the same range as the ED calculation.  The gap can be extracted from the plateau in larger systems as well, although it is evident from the inset in Fig.~\ref{fig:fill_Ncu_2} that the gap opens at significantly lower temperatures at larger $N$.  When $N=16$, the zero-temperature extrapolated (indirect) gap is $\sim0.77$ eV, only about half the size of that at $N=4$.  When $N=36$, the DQMC simulations cannot access sufficiently low temperatures for a gap to fully develop.  Recent experiments have suggested that the indirect gap may be smaller than previously appreciated\cite{Xiang_PRB_2009} (on the order of 0.8 eV).  The agreement with the indirect gap from the larger DQMC system implies that the ED and CPT values may be strongly increased by finite-size effects.

\begin{figure}[t!]
	\includegraphics[width=\columnwidth]{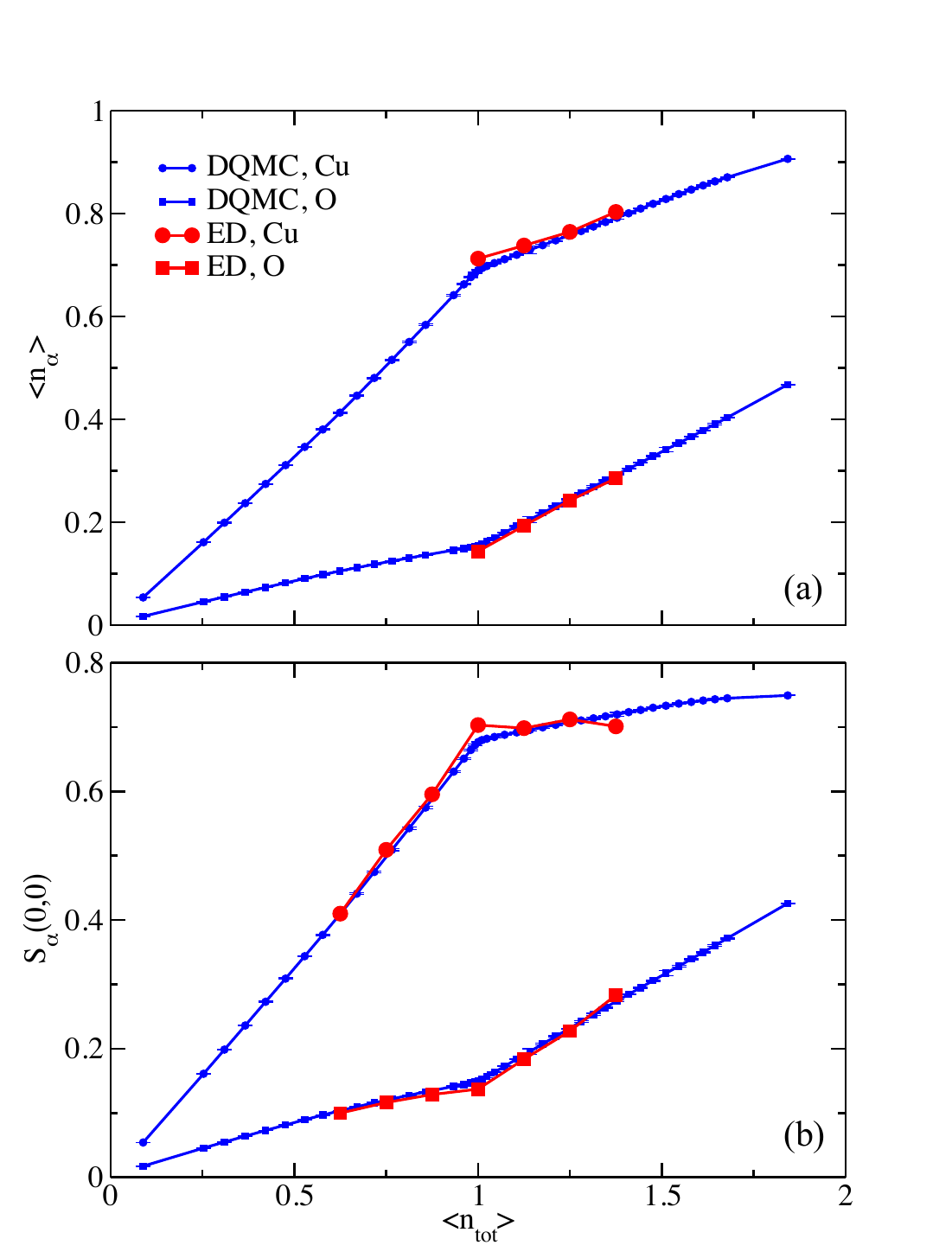}
	\caption{Comparison of the (a) orbitally-resolved filling and (b) local moments on copper and oxygen, showing good quantitative agreement between DQMC and ED.  The DQMC simulations are performed with $N=16$ and $\beta=8$ eV$^{-1}$.}
	\label{fig:DQMC-ED}
\end{figure}

CPT achieves fine momentum resolution at the cost of increased computational complexity associated with the effective Hilbert space size for clusters with open boundary conditions.  As a result, it is much more difficult for CPT to access doped systems, but doping evolution can be compared between DQMC and ED.  DQMC, which works in the grand canonical ensemble, has a continuously tunable filling, while ED is restricted to discrete fillings, but a comparison demonstrates that the overall agreement persists over a wide doping range.  Qualitatively, the spectral functions show a similar degree of agreement at $12.5\%$, $25\%$, and $37.5\%$ hole doping as at $0\%$ doping.  Whereas the identification and interpretation of the direct and indirect gaps are more challenging, as seen above, equal-time quantities provide a clear-cut way of comparing DQMC and ED.  The orbitally-resolved filling curves $\langle \mathrm{n_{Cu,O}} \rangle$ are numerically the same within error bars at many dopings [Fig.~\ref{fig:DQMC-ED}(a)].  The local spin moments also agree well at a quantitative level, and any slight discrepancy can be explained by thermally driven fluctuations of holes between copper and oxygen orbitals [Fig.~\ref{fig:DQMC-ED}(b)].  Hence Fig.~\ref{fig:DQMC-ED} shows that the DQMC and ED results agree well with each other.

These comparisons between DQMC, ED, and CPT calculations demonstrate that despite being limited to high temperature, conclusions drawn from DQMC agree with those from the zero-temperature methods, ED and CPT.  In addition, ED, which is limited to small system sizes, produces results in good agreement with DQMC even on a quantitative level, showing that they access similar physics in spite of temperature and size limitations.  Thus the strengths of ED and CPT can be used to complement those of DQMC to form a more complete picture of the properties of the three-orbital Hubbard model.

\section{Conclusions}
We have characterized the three-orbital Hubbard model using DQMC for a set of parameters applicable to the cuprates.  A systematic exploration of the average fermion sign maps out the effects of different parameters, showing that increasing system size, decreasing temperature, and finite oxygen on-site interactions and oxygen-oxygen hopping reduce the average sign.  At certain doping levels, a coincidence of system geometry leads to local maxima, which may enable better access to intermediate dopings where interesting physics occurs in the cuprates.    

DQMC simulations are used to compute various equal- and unequal-time quantities.  For completeness, the doping dependences of the potential and kinetic energies are explored, demonstrating that the potential energy is governed by the double occupancy, as expected, and that the kinetic energy is controlled by competing tendencies.  As expected for the three-orbital Hubbard model, a plateau corresponding to the Mott gap develops in the filling versus chemical potential curves as temperature decreases.  Doped holes preferentially reside on oxygen orbitals, confirming that the model captures behavior seen in cuprate experiments.  In addition, the copper spin-spin correlation function illustrates the dominance of a tendency towards $(\pi,\pi)$ spin fluctuations near $0\%$ doping and shows how the system crosses over from short-range antiferromagnetic to ferromagnetic correlations in the overdoped system.  However, the density-density correlation function shows no signs of charge order.  The orbitally-resolved spectral functions and DOS add dynamical information and provide a way to make connection with PES experiments.  The doping evolution from the undoped to the overdoped system is explored in detail and demonstrates that inclusion of oxygen orbitals is crucial to capturing the spectroscopic details at higher energies.  

DQMC simulations are compared with the complementary techniques of ED and CPT to form a more complete picture of the three-orbital Hubbard model.  Despite significant differences in system size and temperature, the spectral functions and DOS agree on a qualitative level.  The doping dependence of the orbitally-resolved fillings and local spin moments computed by DQMC and ED also show close agreement.  The charge transfer gap enables a direct comparison of the three methods; the indirect gap from DQMC is found to be significantly smaller than that from ED and CPT.  Using the Mott plateau in the  filling as a proxy for the gap points out a subtlety in interpreting this discrepancy: the interpolated zero-temperature direct gap from DQMC is similar to the direct gap from ED and experiment, while the interpolated indirect gap from DQMC agrees with recent measurements.  Hence, using equal-time quantities, DQMC simulations reveal the difference between the direct and indirect gaps that is hinted at experimentally.

As final note, these calculations reveal the subtlety in using the single-band versus three-orbital Hubbard model to describe the cuprates.  The equal-time quantities such as the filling and spin-spin and density-density correlations remain qualitatively similar whether oxygen on-site interactions are included or neglected, suggesting that oxygen correlations are not significant in energy-integrated measurements and that the single-band description should be adequate.  However, the spectral functions demonstrate the importance of including oxygen orbitals when examining dynamical quantities.  The single-band model cannot capture the evolution of orbital content across the Brillouin zone, which helps to identify the ZRS as well as the distinct orbital characters at the node [$(\pi,\pi)$] and antinode [$(\pi,0)$].  In addition, the three-orbital model includes the true LHB, while the single-band model treats the ZRS band as the LHB and neglects bands further away from the Fermi level.  While the difference between the single-band and three-orbital spectral functions is not significant near the Fermi level, as evidenced by the usefulness of the single-band Hubbard model in describing the cuprates, studying the system at energies away from the Fermi level for comparison to experimental techniques such as PES requires including the oxygen orbitals.  Hence the three-orbital Hubbard model, although more computationally intensive, provides important information on the interplay between different degrees of freedom and adds to our understanding of the cuprates.

This research was supported by the U.S. Department of Energy (DOE), Office of Basic Energy Sciences, Division of Materials Sciences and Engineering, under Contract No. DE-AC02-76SF00515, SLAC National Accelerator Laboratory (SLAC), Stanford Institute for Materials and Energy Sciences. Y.F.K. was supported by the Department of Defense (DOD) through the National Defense Science and Engineering Graduate Fellowship (NDSEG) Program and by the National Science Foundation (NSF) Graduate Research Fellowship under Grant No. 1147470.  C.C.C. is supported by the Aneesur Rahman Postdoctoral Fellowship  at  Argonne  National  Laboratory (ANL),  operated by the US Department of Energy (DOE) Contract No. DE-AC02-06CH11357.  Y.W. was supported by the Stanford Graduate Fellows in Science and Engineering.  E.A.N. also acknowledges support from DOE Er-046169.  S.J. is funded by the University of TennesseeÕs Science Alliance JDRD program; a collaboration with Oak Ridge National Laboratory.  R.T.S. was supported by the Stewardship Science Academic Alliances (SSAA) program of the National Nuclear Security Administration (NNSA).  The computational work was partially performed at the National Energy Research Scientific Computing Center (NERSC), supported by the U.S. DOE under Contract No. DE-AC02-05CH11231.

\bibliography{3band_DQMC_Bib}

\end{document}